\documentclass[aps,pra,reprint,superscriptaddress,showpacs]{revtex4-1}

\usepackage[]{hyperref} 
\usepackage[cp1250]{inputenc}
\usepackage{amsmath}
\usepackage{amsfonts}
\usepackage{textcomp}
\usepackage[]{graphicx}
\usepackage[]{epstopdf} 
\usepackage{mathrsfs}
\usepackage{amsbsy}
\usepackage{cleveref}
\usepackage[rightcaption]{sidecap}
\usepackage[dep]{snapshot} 
\usepackage{xr}   
\pdfoutput=1
\DeclareMathOperator{\sign}{sgn}
\DeclareMathOperator{\di}{d\kern-0.4ex}

\hypersetup{pdfauthor={Wiktor Walasik, Fangwei Ye, Gilles Renversez ^*},pdftitle={Plasmon-soliton waves in planar slot waveguides: II. Results for stationary waves and stability analysis},pdfkeywords={slot waveguide, plasmonics, nonlinear optics, spatial soliton, surface plasmon, stability analysis, bifurcation,  FDTD, stationary states, temporal evolution}}

\begin{document}

\title{Plasmon--soliton waves in planar slot waveguides: \\II. Results for stationary waves and stability analysis}

\author{Wiktor Walasik}
\affiliation{Aix--Marseille Universit\'{e}, CNRS, Centrale Marseille, Institut Fresnel, UMR 7249, 13013 Marseille, France}
\email[corresponding author: ]{gilles.renversez@fresnel.fr}
\affiliation{ICFO --- Institut de Ci\`{e}ncies Fot\`{o}niques, Universitat Polit\`{e}cnica de
Catalunya, 08860 Castelldefels (Barcelona), Spain}
\author{Gilles Renversez} 
\affiliation{Aix--Marseille Universit\'{e}, CNRS, Centrale Marseille, Institut Fresnel, UMR 7249, 13013 Marseille, France}
\author{Fangwei Ye}
\affiliation{Department of Physics and Astronomy, Shanghai Jiao Tong University, Shanghai  200240, China}

\date{\today}

\begin{abstract}
We describe  the results of the two methods we developed to calculate the stationary nonlinear solutions in one-dimensional  plasmonic slot waveguides made of a finite-thickness nonlinear dielectric core surrounded by metal regions. These two methods are described in detail in the preceding article [Walasik et al., submitted to PRA]. 
For symmetric waveguides, we provide the nonlinear dispersion curves obtained using the two methods  and compare them. We describe the well known low-order modes and the higher-modes that were not described before. All the modes are classified into two families: modes with and without nodes. We also compare nonlinear modes with nodes with the linear modes in similar linear slot waveguides with a homogeneous core. We recover the symmetry breaking Hopf bifurcation of the first symmetric nonlinear mode toward an asymmetric mode and we show that one of the higher modes also exhibits a bifurcation. 
We study the behavior of the bifurcation of the fundamental mode as a function of the permittivities of a metal and a nonlinear core. We demonstrate that the bifurcation can be obtained at low power levels in structures with optimized parameters.
Moreover, we provide the dispersion curves for asymmetric nonlinear slot waveguides.  Finally, we give results concerning the stability of the fundamental symmetric mode and the asymmetric mode that bifurcates from it using both theoretical argument and numerical propagation simulations from two different full-vector methods. We investigate also the stability properties of the first antisymmetric mode using our two numerical propagation methods.
\end{abstract}

\pacs{42.65.Wi, 42.65.Tg, 42.65.Hw, 73.20.Mf}
\keywords{Nonlinear waveguides, Optical solitons, Kerr effect: nonlinear optics, Plasmons on surfaces and interfaces / surface plasmons}

\maketitle


\section{Introduction}

Nonlinear slot waveguides (NSWs) are structures made of a finite-size nonlinear dielectric layer sandwiched between two semi-infinite metal layers.
They have been studied in Refs.~\cite{Feigenbaum07,Davoyan09a} where it was shown that they allow for sub-wavelength confinement of light and phase matching for the second-harmonic generation.  
More recently, in the works of Rukhlenko \textit{et al.}~\cite{Rukhlenko11,Rukhlenko11a}, analytical formulas for the dispersion relations of these NSWs were presented for symmetric and antisymmetric nonlinear modes only. These dispersion relations were given using integral equations that have to be solved numerically. The study of Davoyan \textit{et al.}~\cite{Davoyan08} showed, using the numerical shooting method to solve Maxwell's equation in NSWs, that a symmetry breaking bifurcation that generates an asymmetric mode from the fundamental symmetric mode occurs in NSWs. 
Such bifurcation phenomena are well known in fully dielectric nonlinear structures~\cite{Langbein_85II,Akhmediev82,Moloney86,Chiang93,Chiang94,Boardman86a,Holland86,Sukhorukov01}. 
Recently, higher order modes were also reported in NSWs~\cite{Walasik14a}. 
Moreover, it was shown that plasmonic coupling and symmetry breaking phenomena can be observed in waveguides built of a linear dielectric core sandwiched by nonlinear metals~\cite{Davoyan11,Davoyan11a}.
Nonlinear switching was predicted in NSW-based structures using numerical simulations~\cite{Nozhat12}.

In the preceding article~\cite{Walasik14b}, we describe two models we developed to study the stationary nonlinear solutions in NSWs where the nonlinear core of the focusing Kerr type was considered. 
The first model assumes that the nonlinear term depends only on the transverse component of the electric field and that the nonlinear refractive index change is small compared to the linear part of the refractive index. It allows us to describe analytically the field profiles in the whole waveguide. It also provides a closed analytical formula for the nonlinear dispersion relation. This first model is called Jacobi elliptic function based model (JEM).
The second model takes into account the full dependency of the Kerr nonlinear term on all electric field components and no assumption is required on the amplitude of the nonlinear term. The disadvantage of this approach is the fact that the field profiles must be computed numerically. This second model is called the interface model (IM).

This article is organized in the following way.
In Sec.~\ref{sec:res-sym}, we describe the results obtained with our two models for symmetric NSWs. They include a mode classification taking into account the higher order modes we found previously~\cite{Walasik14a} and a detailed study of the field profile transformation as a function of power. We also provide a permittivity contrast study that allows us to decrease by several orders of magnitude the bifurcation threshold at which the first asymmetric mode appears. In Sec.~\ref{sec:res-asym}, we provide the results concerning asymmetric NSWs in which the mode degeneracy is lifted. Finally, in Sec.~\ref{res-stability}, using both theoretical arguments and numerical propagation simulations from two different full-vector methods we provide results on the stability of the main stationary solutions obtained in symmetric NSWs. 

\section{Results for symmetric waveguides}
\label{sec:res-sym}

\subsection{Dispersion relations, field profiles and mode classification}
\label{sec:res-sym-disp}

In this section, the dispersion relations obtained for the symmetric NSW are presented. The field profiles corresponding to each of the dispersion curves are depicted and allow us to classify the modes according to their symmetry and the number of nodes in the magnetic field profile. 

\begin{figure}[!b]
	\centering
	\includegraphics[width=1.0\columnwidth,clip=true,trim= 0 0 0 0]{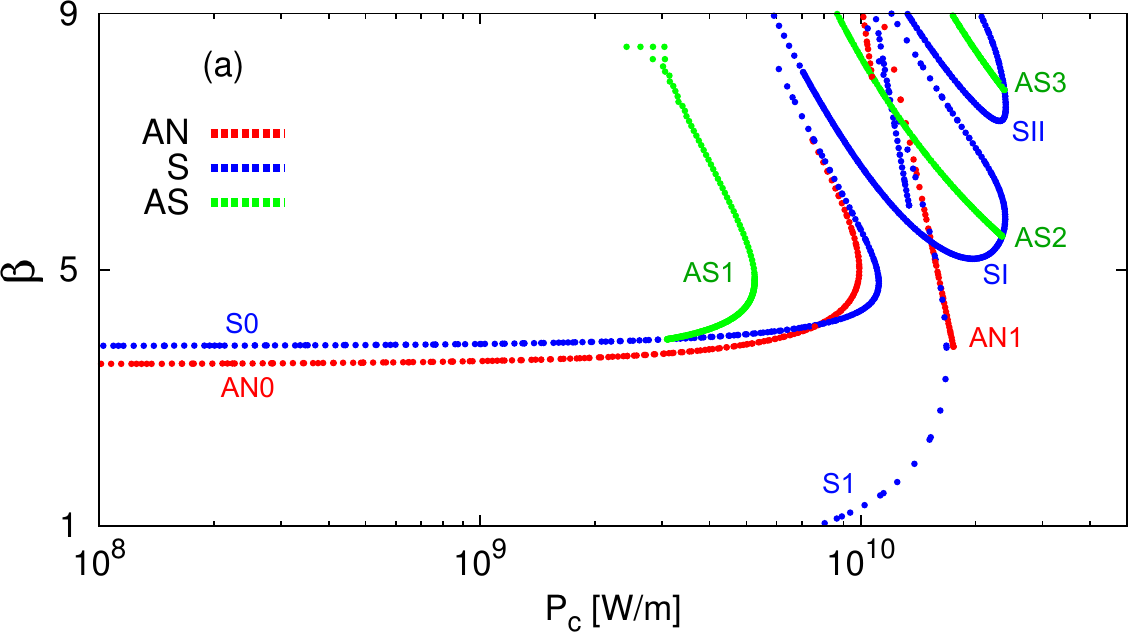}
	\includegraphics[width=1.0\columnwidth,clip=true,trim= 0 0 0 0]{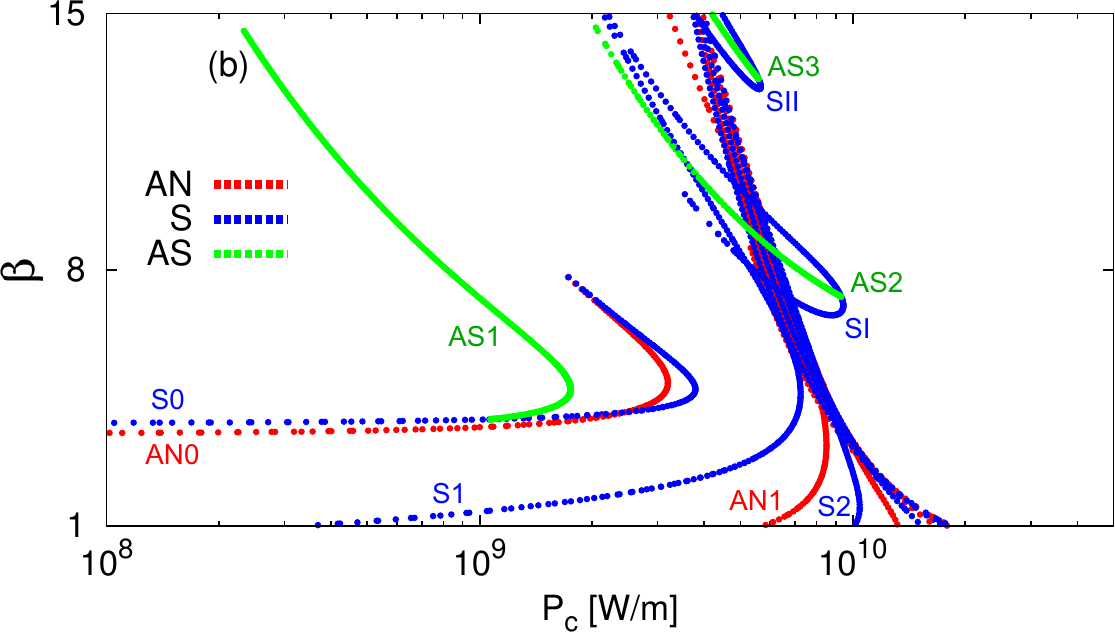}
	\caption{Dispersion diagrams $\beta(P_c)$ for the symmetric NSW obtained using (a) the JEM and (b) the IM.}
	\label{fig:disp-Pc}
\end{figure}

\begin{figure}[!t]
	\centering
	\includegraphics[width=0.95\columnwidth,clip=true,trim= 0 0 0 0]{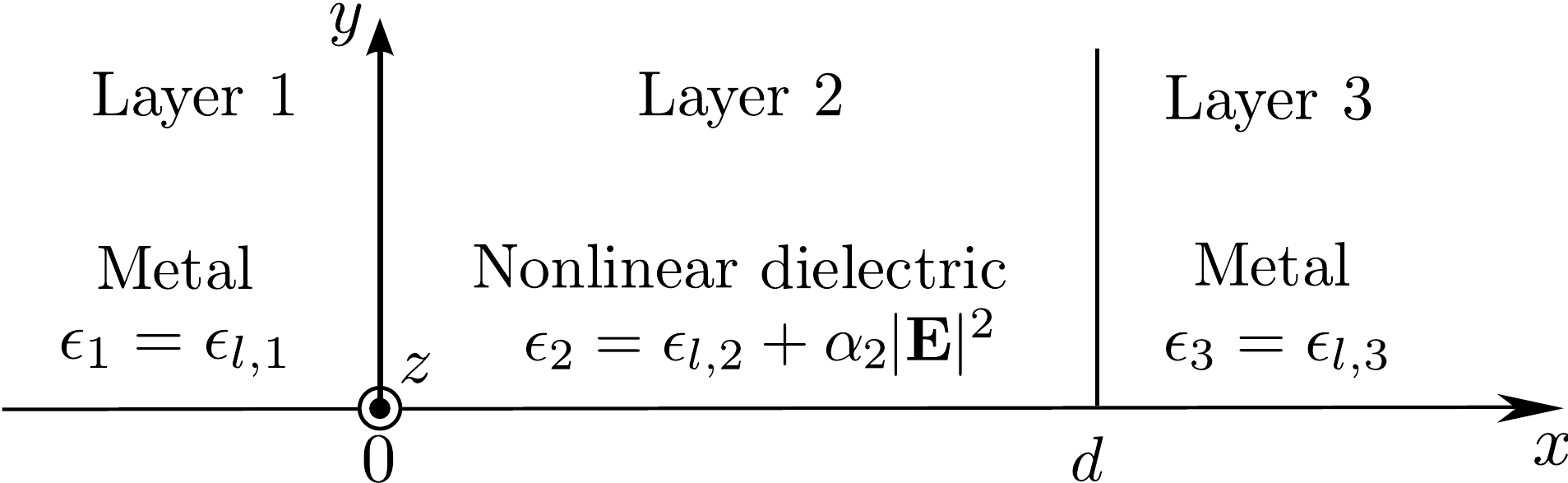}
	\caption{Geometry of the plasmonic NSW with the parameters of the structure.}
	\label{fig:geometry-slot}
\end{figure}

Figure~\ref{fig:disp-Pc} presents dispersion relations for the symmetric NSW obtained using the JEM and the IM. The parameters of the NSW studied are: $\epsilon_1 = \epsilon_3 = -90$ (gold), $\epsilon_{l,2} = 3.46^2$, $\alpha_2 = 6.3\cdot10^{-19}$~m$^2$/V$^2$ (hydrogenated amorphous silicon) and $d = 400$~nm at a wavelength $\lambda = 1.55$~$\mu$m. The geometry of the structure with its parameters is shown in Fig.~\ref{fig:geometry-slot}. The dispersion relations present the dependence of the effective index of the mode~$\beta$ as a function of the power density in the waveguide core~$P_c$ which is calculated in the following way:
\begin{equation}
P_c = \int_{0}^{d} S_z \mathrm{d}x,
\end{equation}
where $S_z$ denotes the $z$-component of the Poynting vector $\textbf{S} = 1/2\Re e(\textbf{E}\times \textbf{H}^*)$.

We observe a very good qualitative agreement between the dispersion diagrams obtained using our two models. The number and the character of the dispersion curves is very similar in both cases. The qualitative agreement between the results of the two models confirms their validity. Quantitatively speaking, the models agree in the range of low power densities (below $10^9$~W/m). Above this value we observe quantitative differences in the results. The origin of the differences is explained by the assumptions made in the JEM (low nonlinearity, only $E_x$ component of the electric field contributing to the Kerr nonlinear effect) as described in Ref.~\cite{Walasik14b}. In the following, we will focus on the results obtained using the more accurate IM.

\begin{figure}[!ht]
	\centerline{\includegraphics[width=0.50\columnwidth,clip=true,trim= 0 0 0 0]{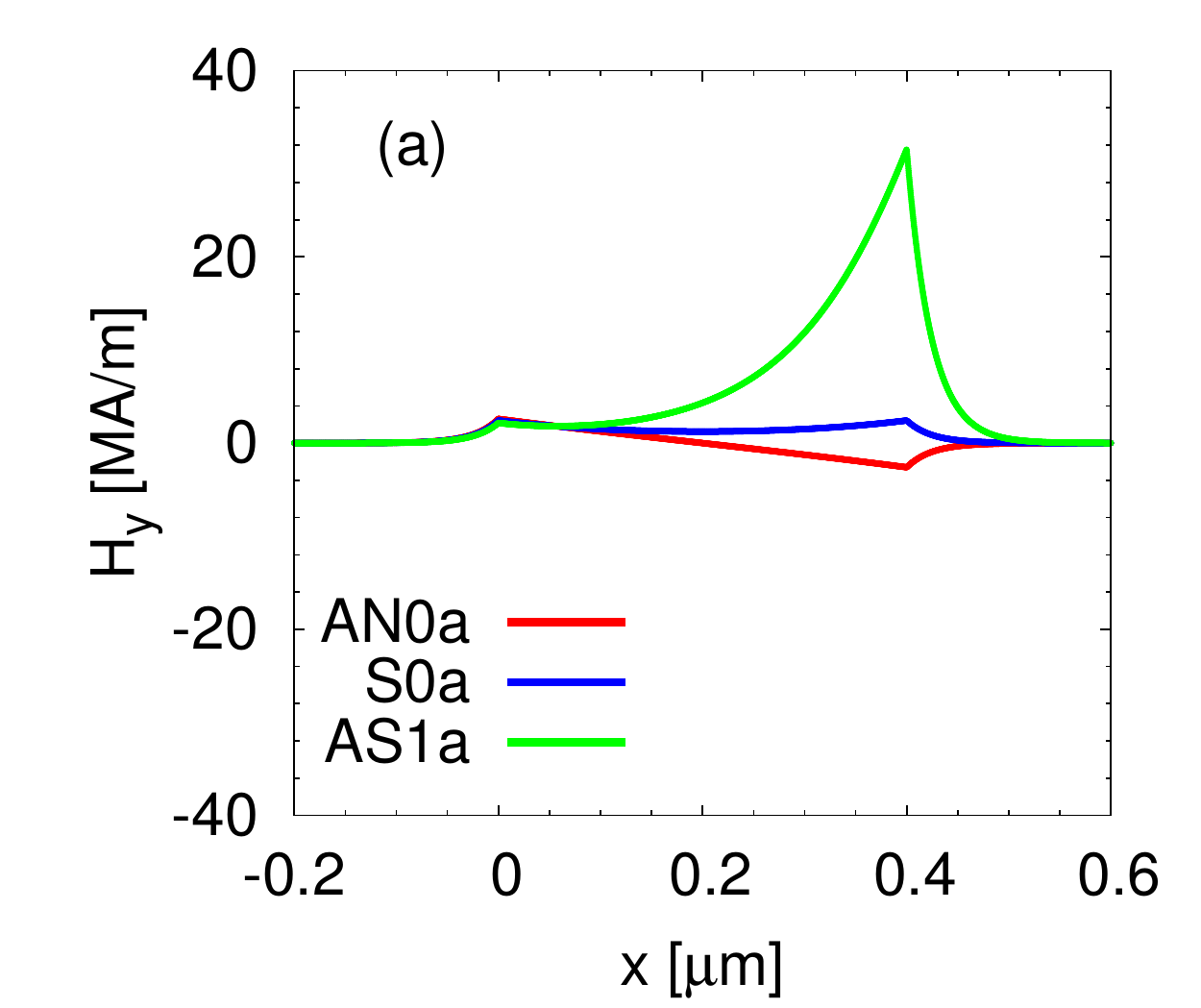}
		\includegraphics[width=0.50\columnwidth,clip=true,trim= 0 0 0 0]{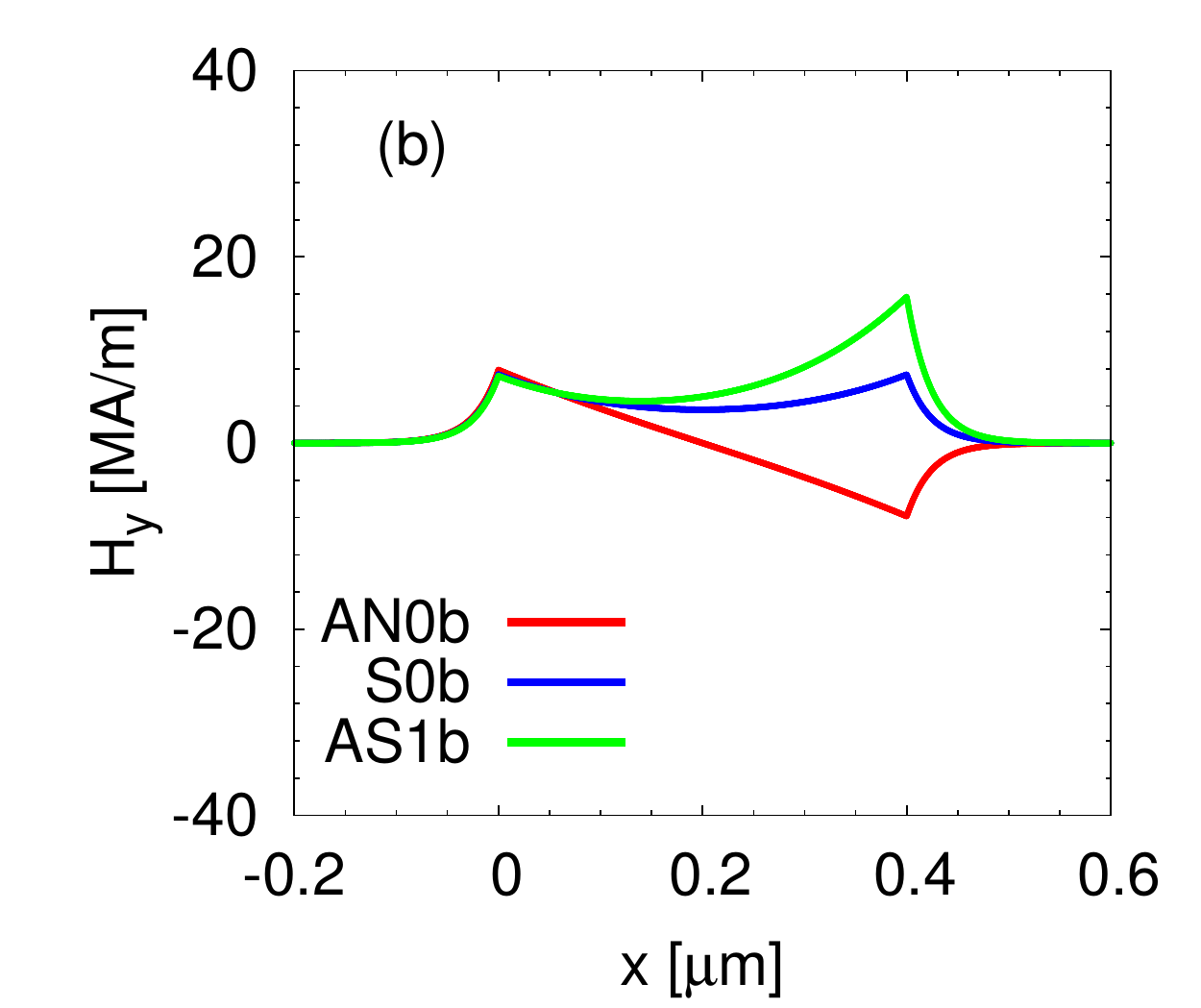}}
	\centerline{
		\includegraphics[width=0.50\columnwidth,clip=true,trim= 0 0 0 0]{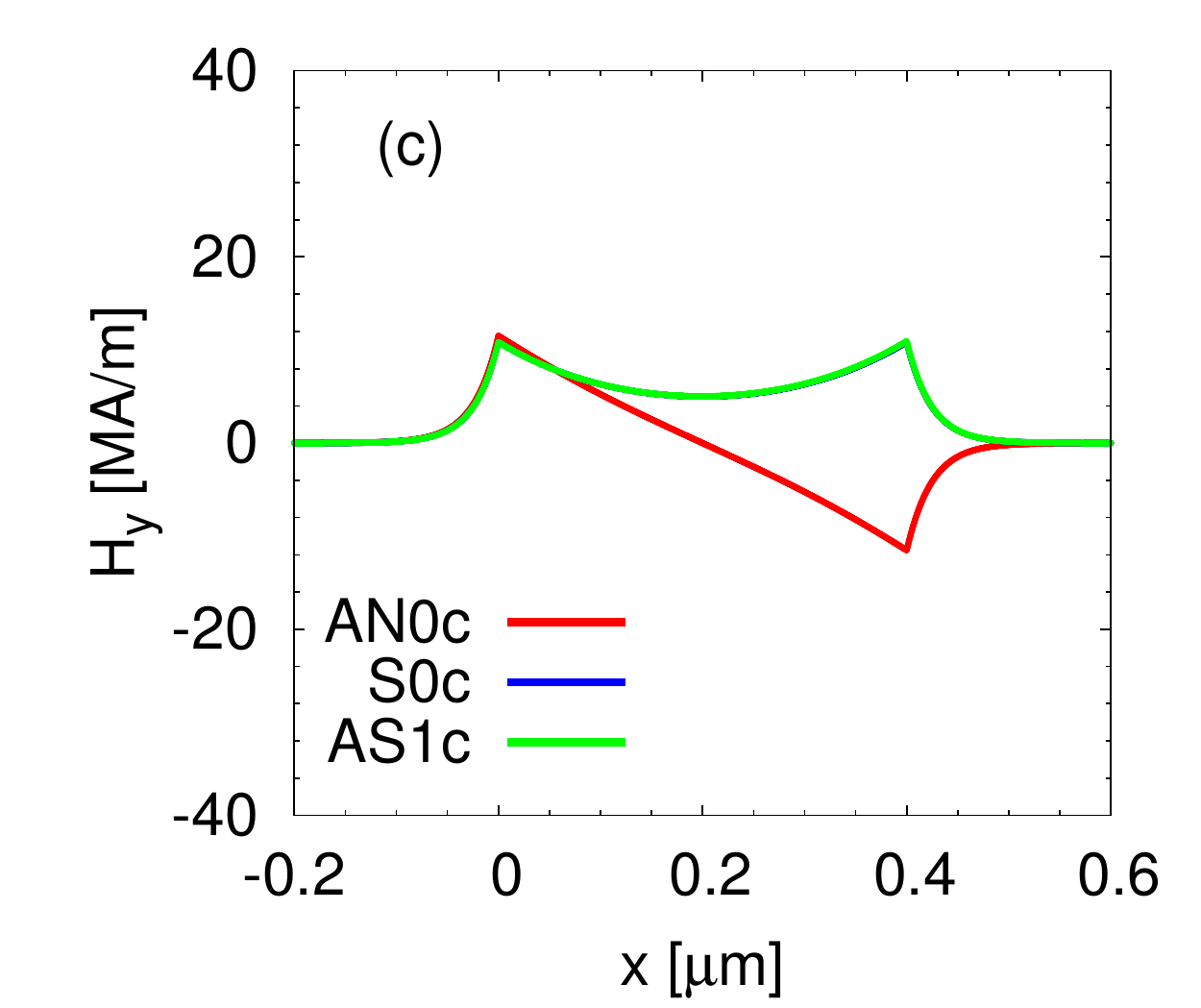}}
	\centerline{\includegraphics[width=0.50\columnwidth,clip=true,trim= 0 0 0 0]{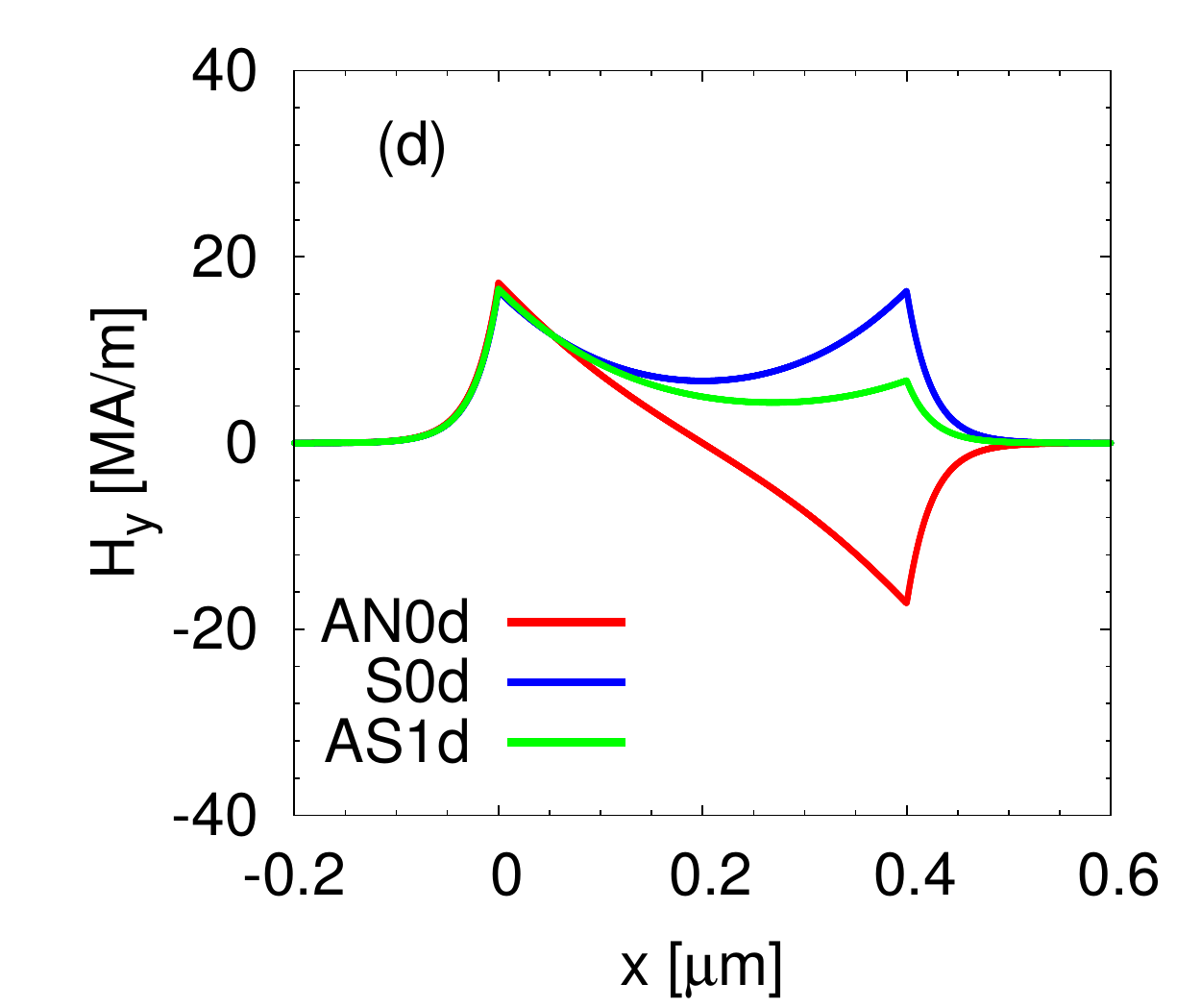}
		\includegraphics[width=0.50\columnwidth,clip=true,trim= 0 0 0 0]{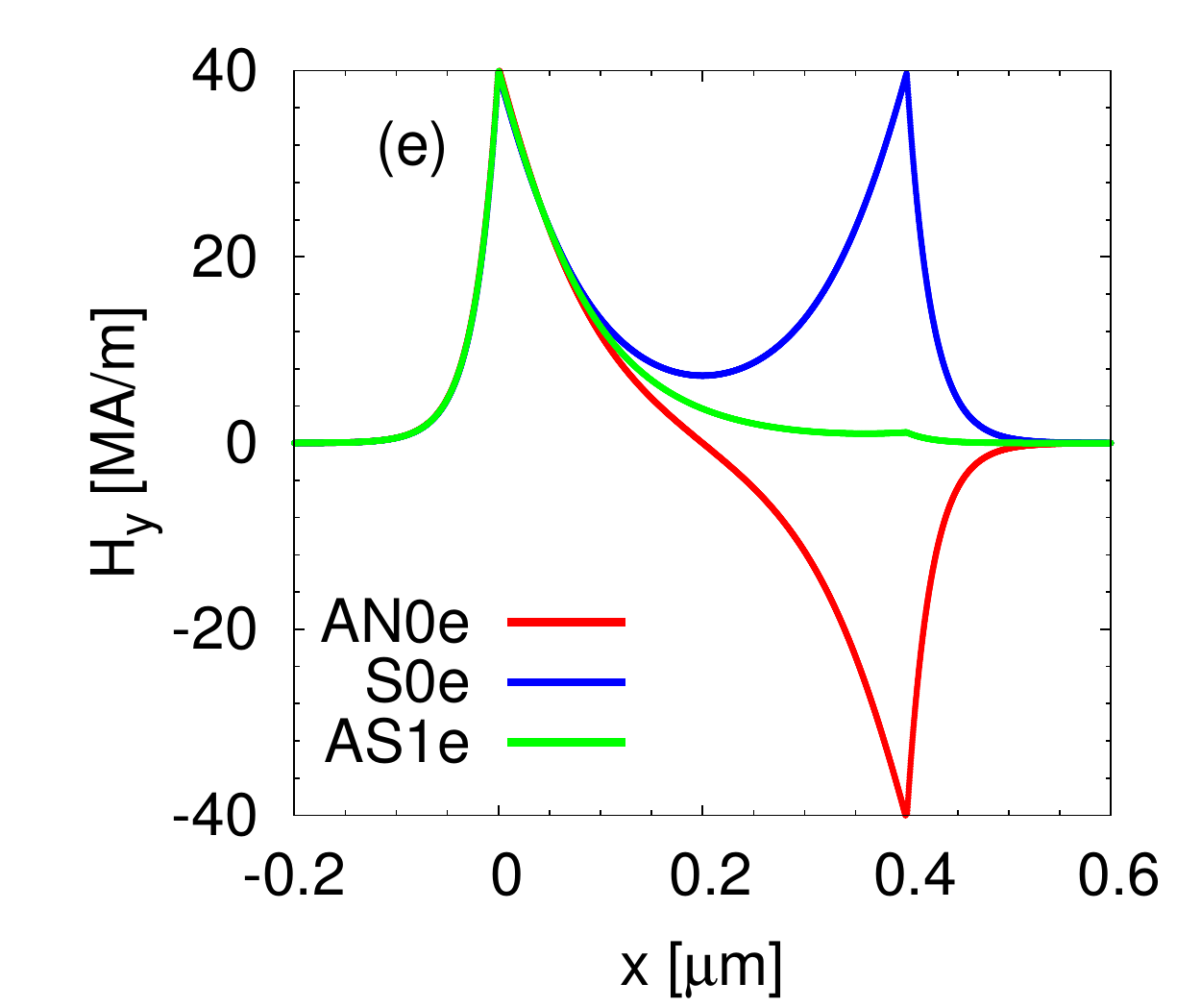}}
	\caption{Profiles of magnetic field $H_y(x)$ for the symmetric S0 mode (blue), antisymmetric AN0 mode (red), and the first-order asymmetric AS1 mode (green). The subplots present the transformation of the field profiles at the points corresponding to the vertical lines labeled a-e indicated in Fig.~\ref{fig:disp-E0}.}
	\label{fig:fields-Hy-sym0-anti0-asym1}
\end{figure}

The NSW supports numerous modes with various properties. Firstly, we will discuss the mode classification according to the symmetry of the mode. For the low power region, the NSW studied here supports two modes: a fundamental symmetric mode [blue curve labeled S0 in Fig.~\ref{fig:disp-Pc} and  blue field profiles in Fig.~\ref{fig:fields-Hy-sym0-anti0-asym1}] and a low-power antisymmetric mode [red curve labeled AN0 in Fig.~\ref{fig:disp-Pc} and  red field profiles in Fig.~\ref{fig:fields-Hy-sym0-anti0-asym1}]. At $P_c\approx 10^9$~W/m a symmetry breaking bifurcation occurs that gives birth to an asymmetric mode~\cite{Davoyan08} [green curve labeled AS1 in Fig.~\ref{fig:disp-Pc} and green field profiles in Fig.~\ref{fig:fields-Hy-sym0-anti0-asym1}]. This modes and this type of behavior are known in nonlinear waveguides~\cite{Davoyan08,Davoyan11,Langbein_85II,Akhmediev82,Moloney86,Chiang93,Chiang94,Boardman86a,Holland86,Sukhorukov01,Akhmediev97}. The power density $P_c$ of the modes S0, AN0, and AS1 first increases with the increase of the effective index $\beta$ and decreases for $\beta \gtrsim 4.75$.

\begin{figure}[!t]
  \centerline{\includegraphics[width=0.50\columnwidth,clip=true,trim= 0 0 0 0]{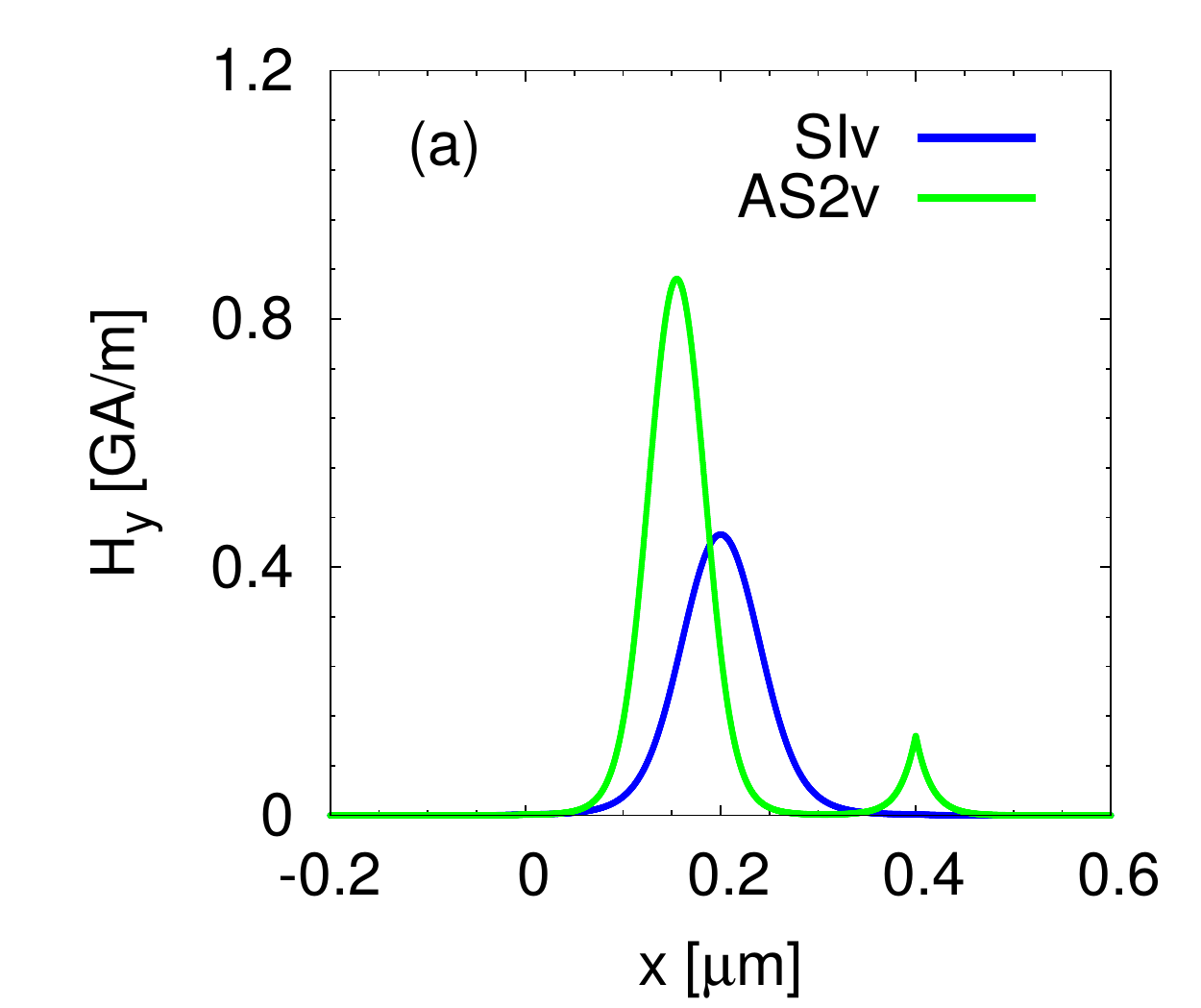}
    \includegraphics[width=0.50\columnwidth,clip=true,trim= 0 0 0 0]{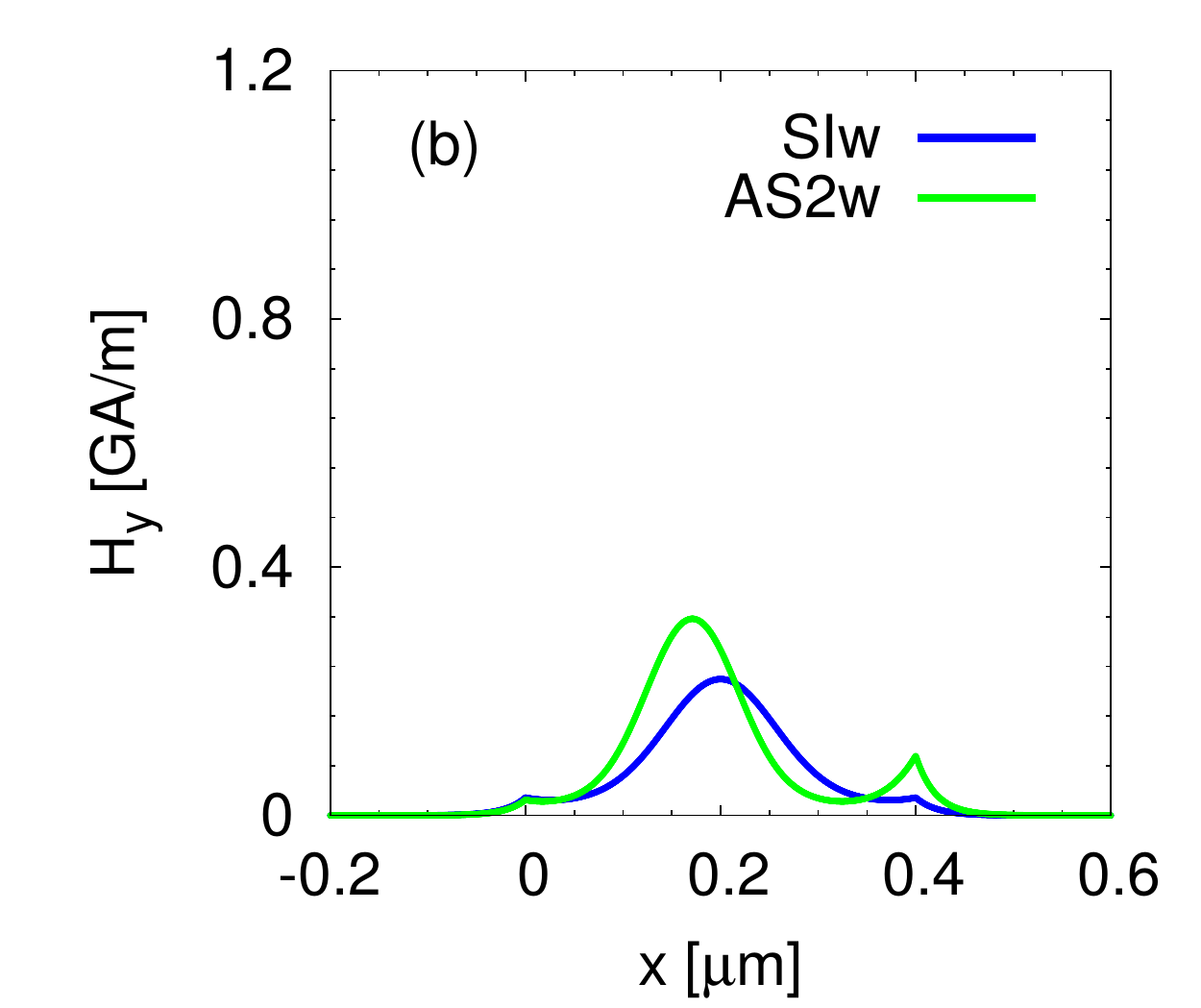}}
  \centerline{
    \includegraphics[width=0.50\columnwidth,clip=true,trim= 0 0 0 0]{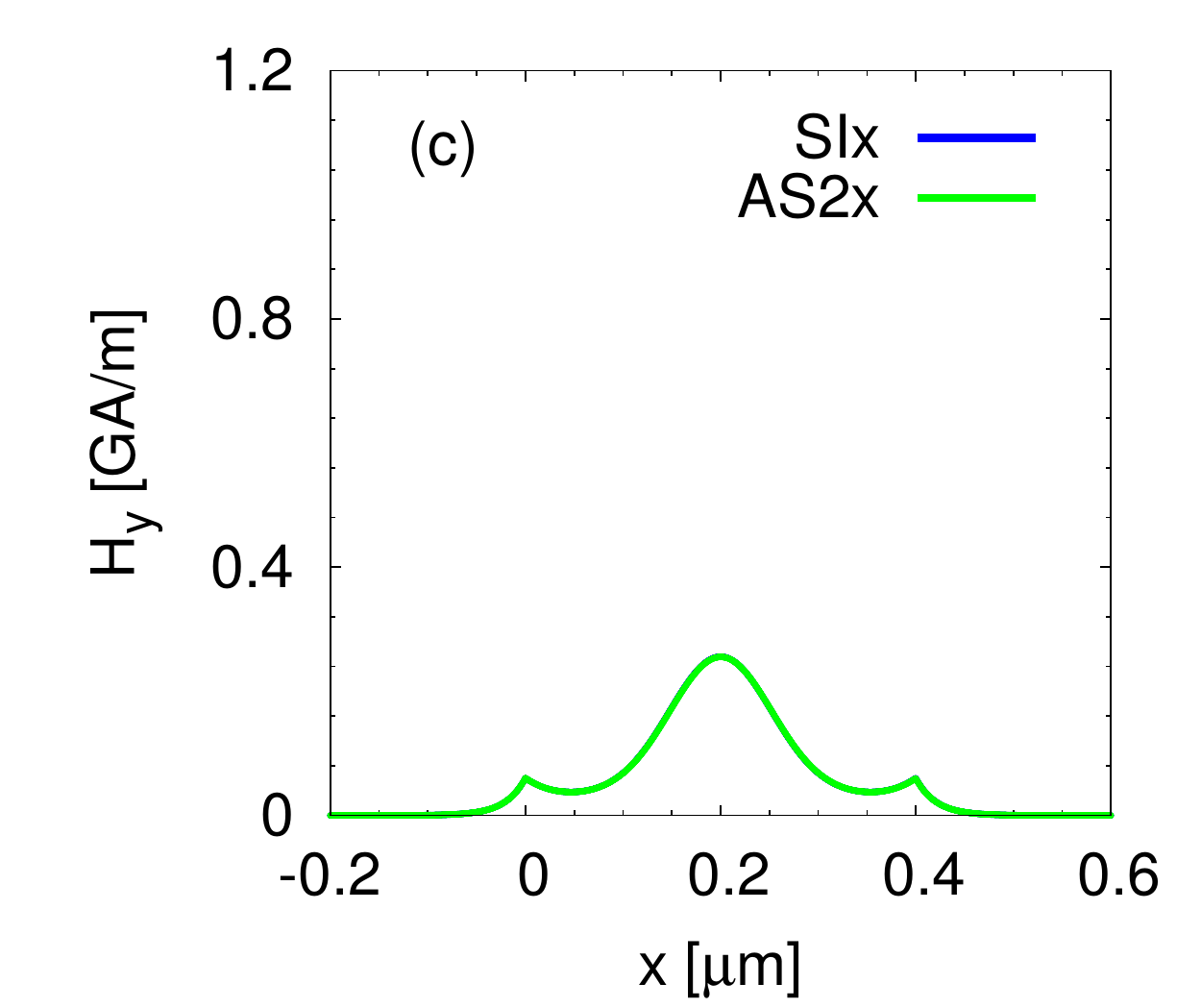}}
  \centerline{\includegraphics[width=0.50\columnwidth,clip=true,trim= 0 0 0 0]{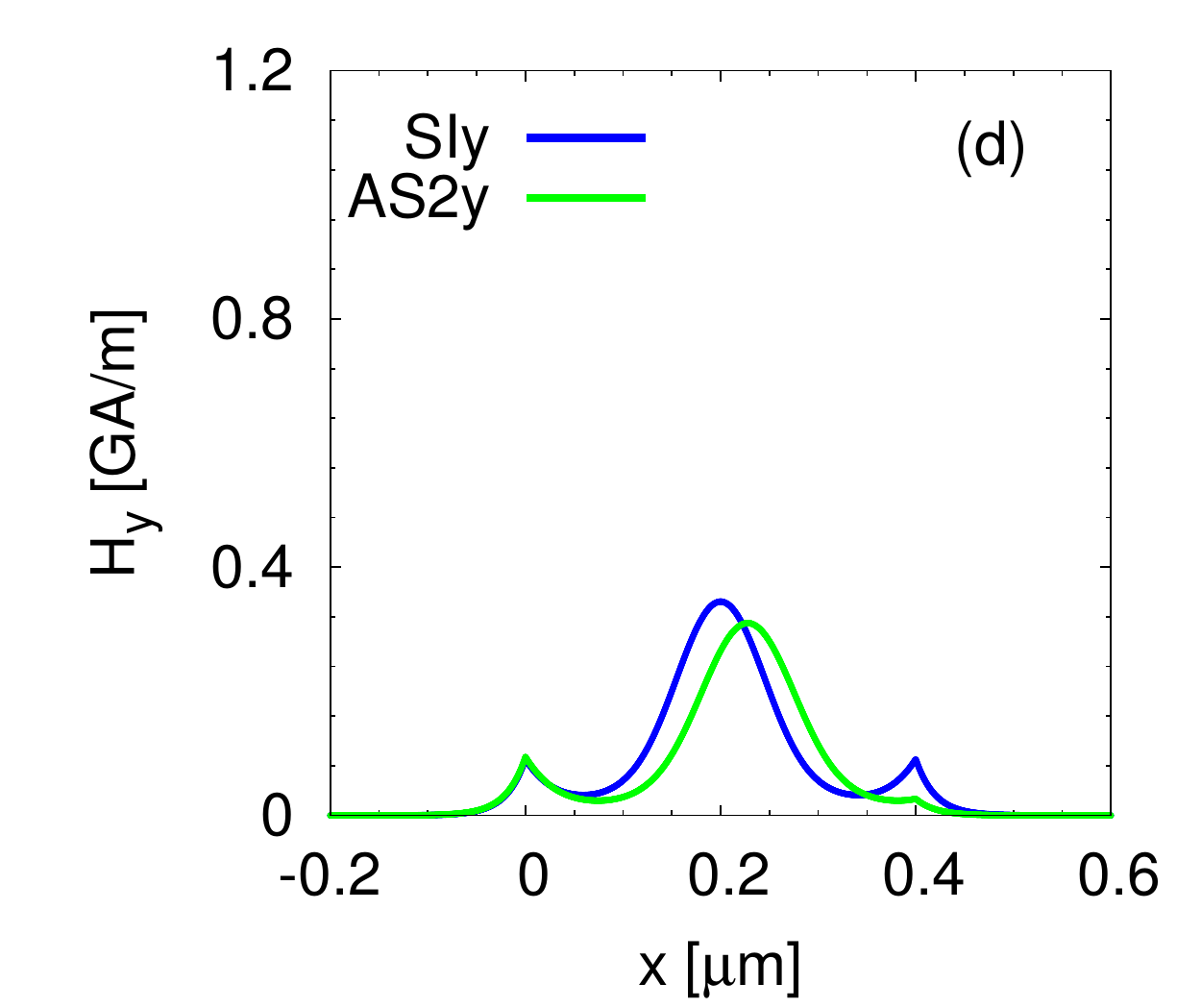}
    \includegraphics[width=0.50\columnwidth,clip=true,trim= 0 0 0 0]{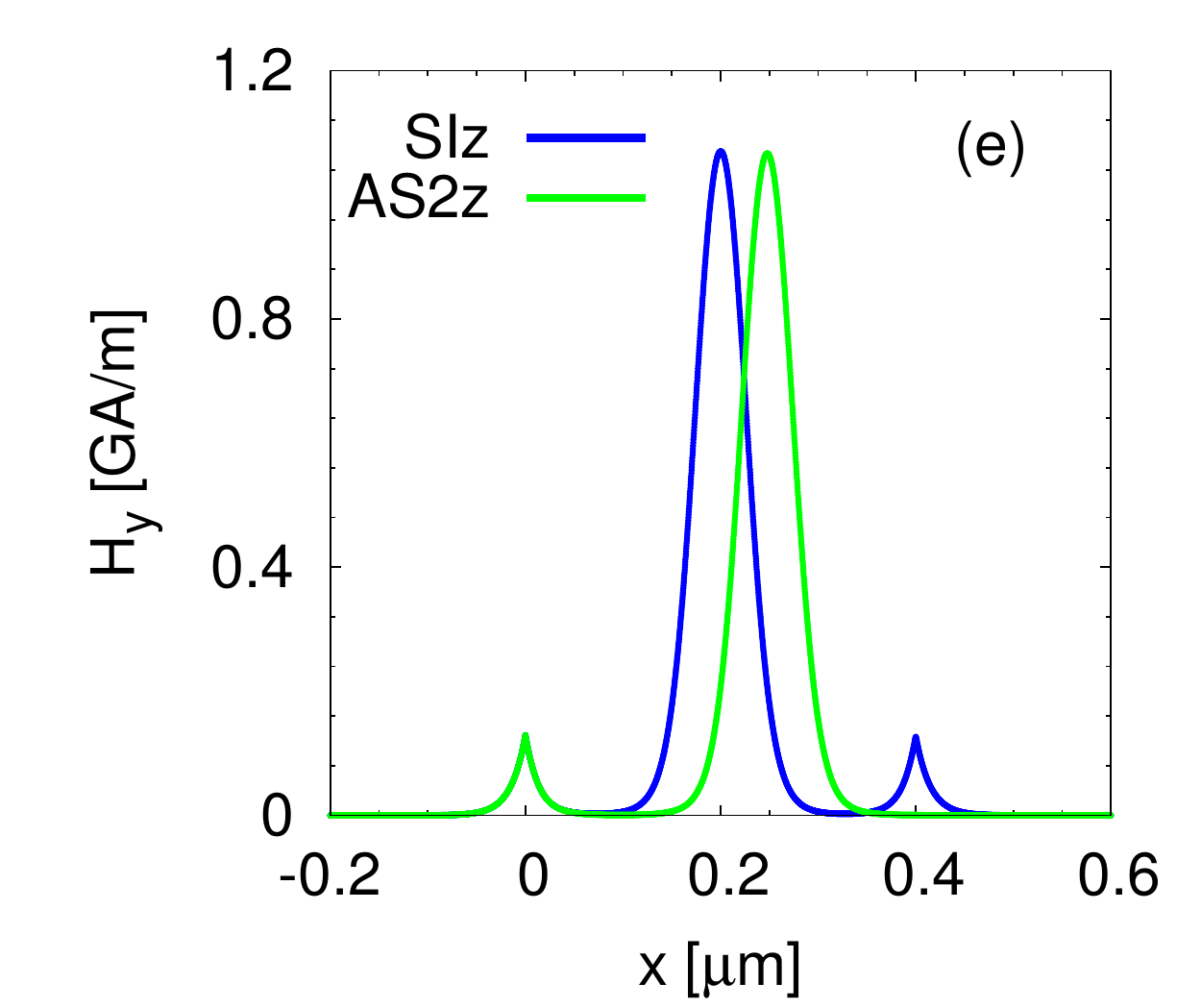}}
  \caption{Profiles of magnetic field $H_y(x)$ for the symmetric SI mode (blue) and the second-order asymmetric AS2 mode (green). The subplots present the transformation of the field profiles at the points corresponding to the vertical lines labeled v-z indicated in Fig.~\ref{fig:disp-E0}. In each subplot the value of $E_0$ \{see Eq.~(\ref{eqn:total-elect}) in Ref.~\cite{Walasik14b}\ for its definition\} is identical for both modes.}
\label{fig:fields-Hy-symI-asym2}
\end{figure}

\begin{figure}[!ht]
	\centerline{\includegraphics[width=0.50\columnwidth,clip=true,trim= 0 0 0 0]{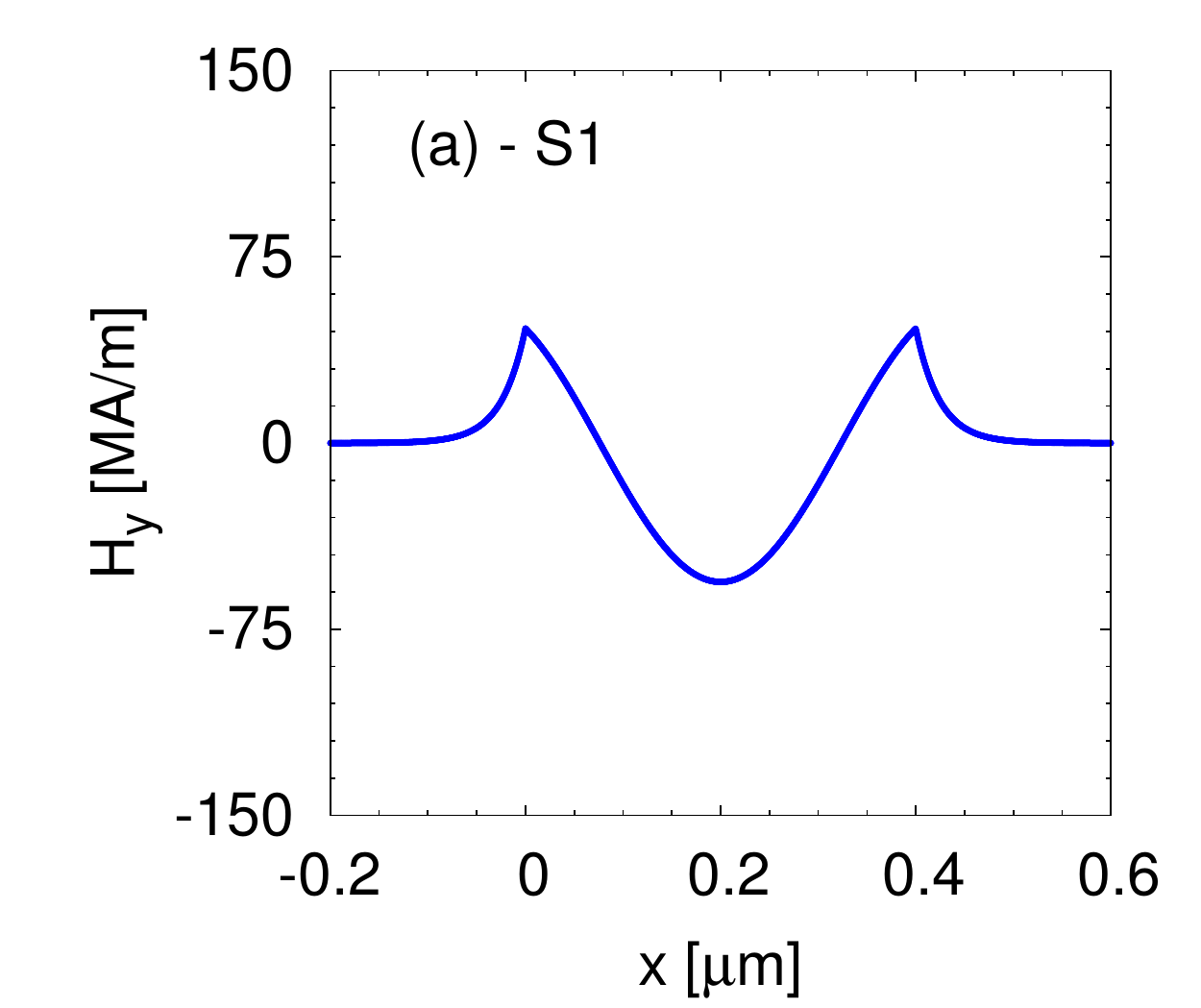}
		\includegraphics[width=0.50\columnwidth,clip=true,trim= 0 0 0 0]{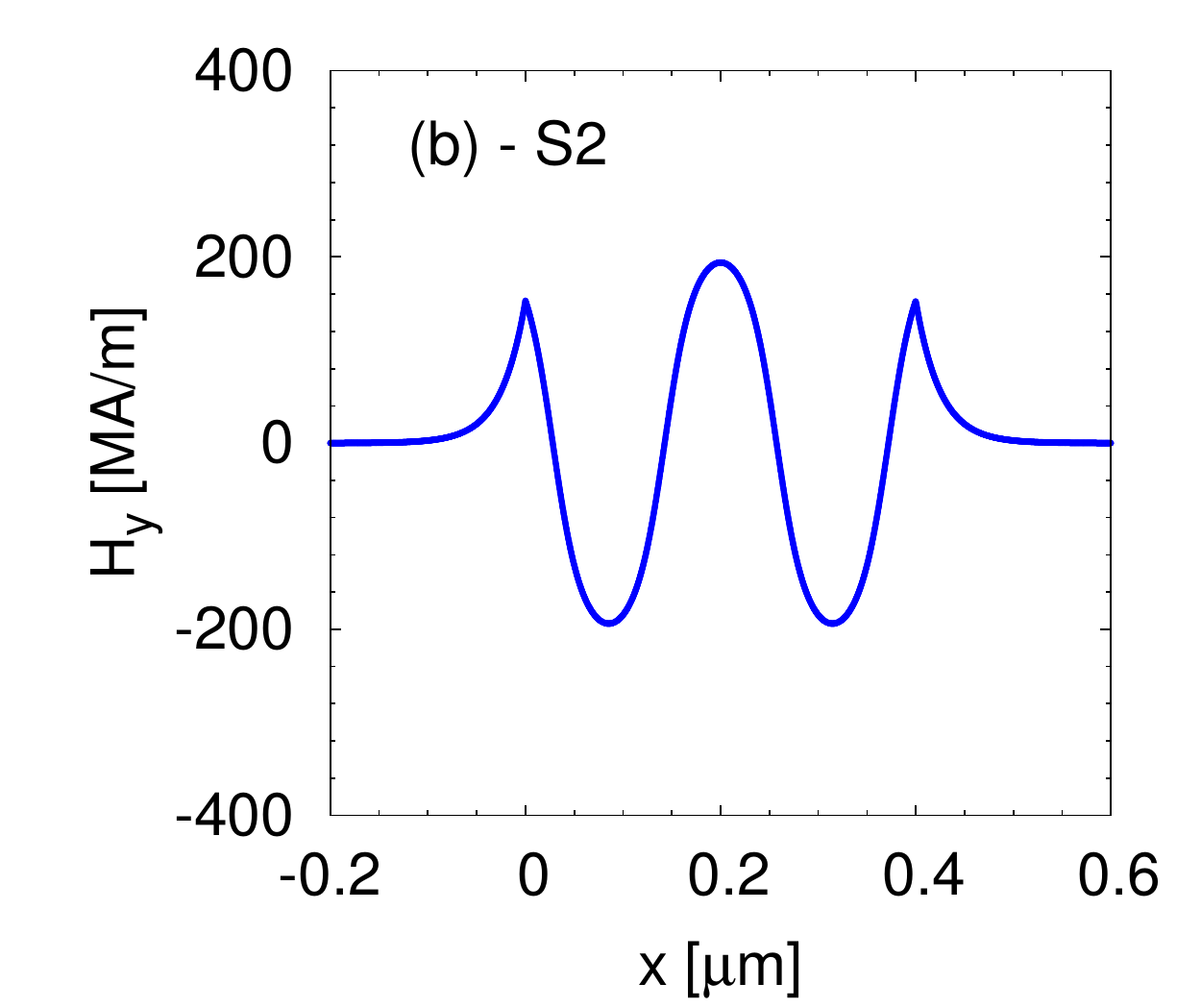}}
	\centerline{\includegraphics[width=0.50\columnwidth,clip=true,trim= 0 0 0 0]{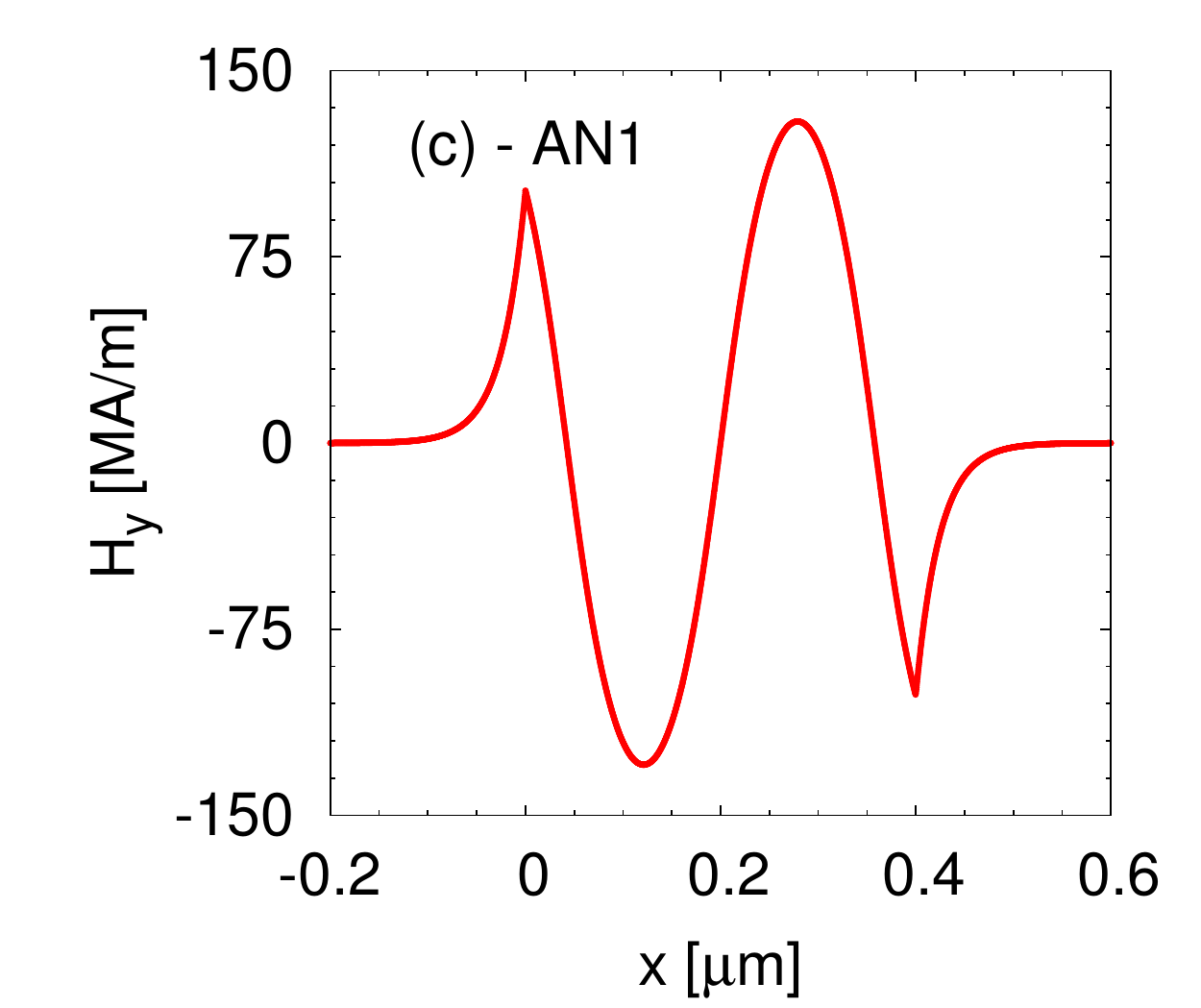}
		\includegraphics[width=0.50\columnwidth,clip=true,trim= 0 0 0 0]{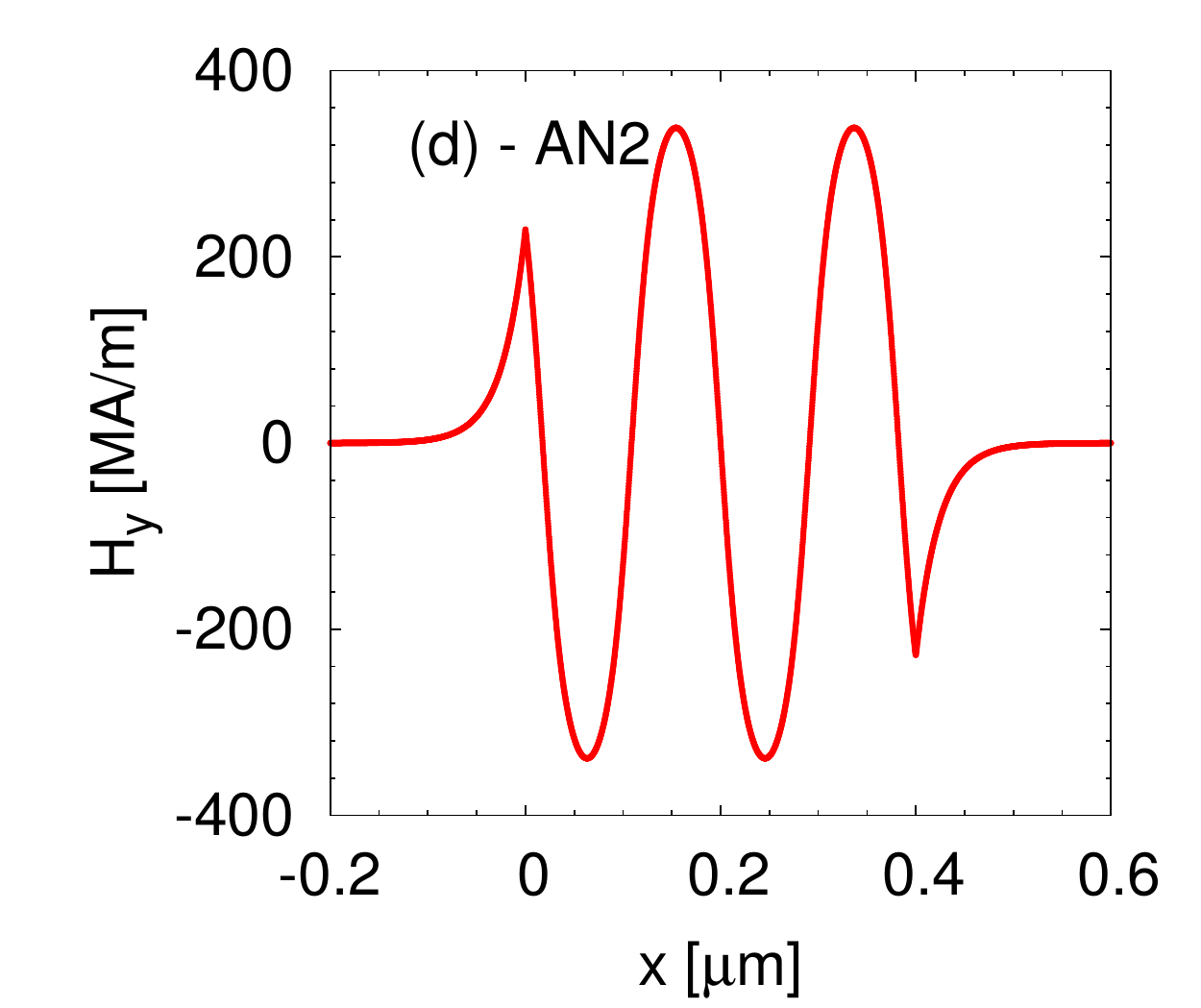}}
	\centerline{\includegraphics[width=0.50\columnwidth,clip=true,trim= 0 0 0 0]{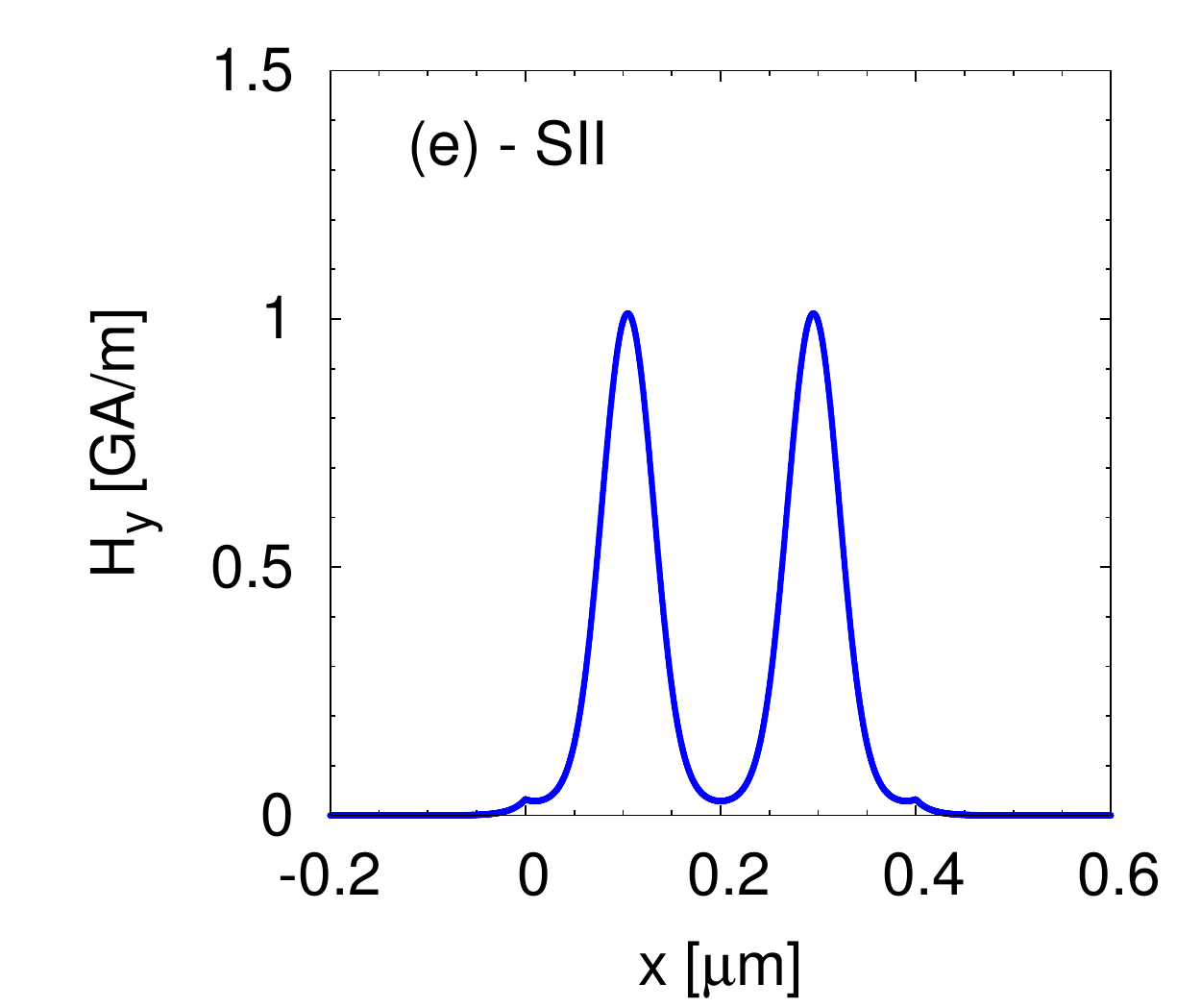}
		\includegraphics[width=0.50\columnwidth,clip=true,trim= 0 0 0 0]{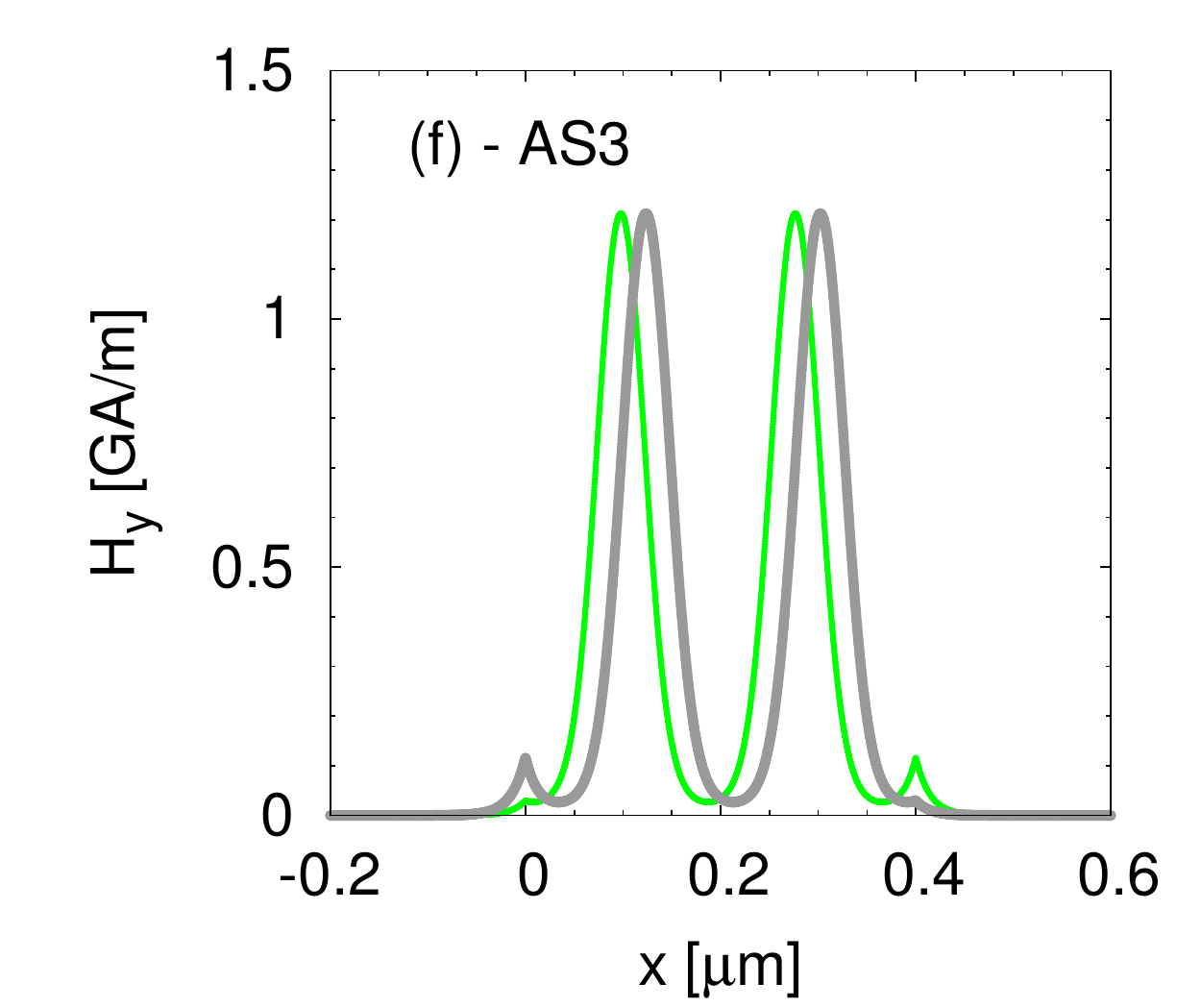}}
	\caption{Typical magnetic field profiles $H_y(x)$, obtained using the IM, corresponding to different dispersion curves indicated in Fig.~\ref{fig:disp-E0}. Abbreviations next to the subfigure labels indicate the dispersion curve to which a given profile corresponds. The color of the profile allows us to distinguish the mode symmetry: symmetric (S --- blue), antisymmetric (AN --- red), and asymmetric (AS --- green). For asymmetric doubly degenerate mode AS3 the second profile is shown in gray.}
	\label{fig:IM-field-examples}
\end{figure}

Our models allow us to find new, higher order modes in NSWs. The higher order modes can be divided into two families: node-less modes and modes with nodes. Among the node-less modes we find higher order symmetric modes (SI and SII) from which asymmetric modes bifurcate (AS2 and AS3, respectively). Their dispersion curves are labeled with the name of the mode in Fig.~\ref{fig:disp-Pc} and their field profiles are presented in Figs.~\ref{fig:fields-Hy-symI-asym2} and \ref{fig:IM-field-examples}(e),~(f). Higher order node-less modes resemble a single soliton (SI and AS2) or two solitons (SII and AS3) propagating in the NSW core.

All the dispersion curves of the asymmetric modes are doubly degenerate. This means that to one value of the effective index (and the corresponding power density) correspond two solutions localized on one of the two core interfaces [compare green curves in  Figs.~\ref{fig:fields-Hy-sym0-anti0-asym1}(a) and (e) for the AS1 mode, Figs.~\ref{fig:fields-Hy-symI-asym2}(a) and (e) for the AS2 mode and the green and the gray profiles in Fig.~\ref{fig:IM-field-examples}(f)]  

The higher order modes with nodes resemble the modes of a linear slot waveguide with a higher refractive index than the one used here (see Sec.~\ref{sec:comp-lin}). Only symmetric (S1, S2, \dots) and antisymmetric (AN1, AN2, \dots) modes with nodes exist. Their dispersion curves are presented in Fig.~\ref{fig:disp-Pc} and their field profiles are shown in Figs.~\ref{fig:IM-field-examples}(a)--(d). The dispersion curves of the modes with nodes start for $\beta = 1$ and their effective index grows with the increase of the power density $P_c$.

\begin{figure*}[!t]
	\centering
	\includegraphics[width=1.0\textwidth,clip=true,trim= 0 0 0 0]{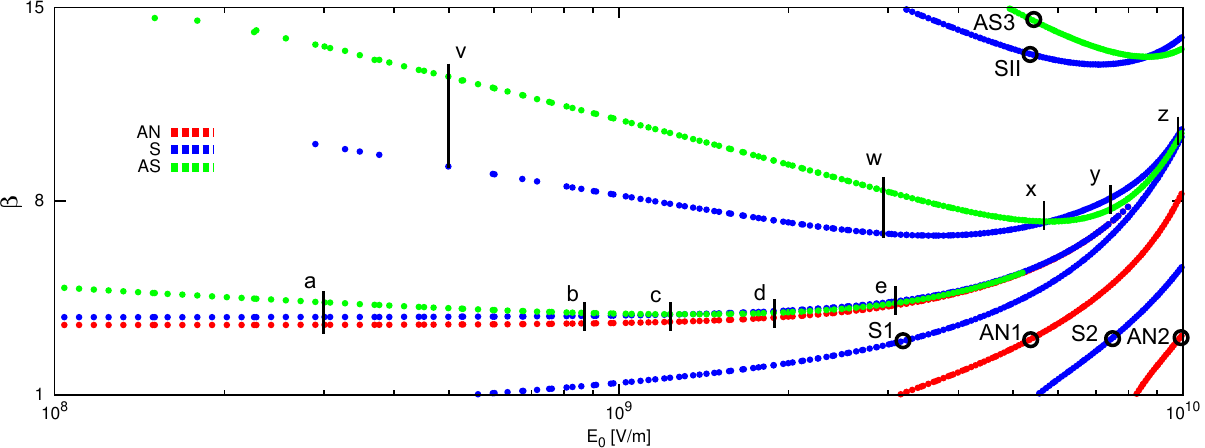}
	\caption{Dispersion diagram $\beta(E_0)$ for the symmetric NSW obtained using the IM. The vertical black lines indicate to the values of the total electric field amplitude $E_0$ corresponding to the field profiles depicted in Figs.~\ref{fig:fields-Hy-sym0-anti0-asym1} and~\ref{fig:fields-Hy-symI-asym2}. Open circles correspond to the field profiles shown in Fig.~\ref{fig:IM-field-examples}.}
	\label{fig:disp-E0}
\end{figure*}

In Fig.~\ref{fig:disp-E0}, we present the dispersion relations obtained using the IM in a different coordinate frame. This time we use the total electric field intensity at $x=0$ (the interface between the NSW core and the metal cladding) $E_0$ \{see Eq.~(\ref{eqn:total-elect}) in Ref.~\cite{Walasik14b}\}. This quantity is one of the input parameters of the IM. The dispersion diagrams $\beta(E_0)$ have a drastically different character from the $\beta(P_c)$ diagrams presented in Fig.~\ref{fig:disp-Pc}. The difference is caused by the fact that $E_0$ is a local quantity, whereas $P_c$ is a global quantity, that results from the integration over the core width. In the coordinates of $E_0$, the dispersion curves of the asymmetric modes are not degenerate. In Fig.~\ref{fig:disp-E0}, we notice that for asymmetric modes, to a given value of $\beta$ correspond two values of $E_0$, that represent solutions localized on the left and right interface of the waveguide core.

In Fig.~\ref{fig:fields-Hy-sym0-anti0-asym1}, the comparison of the field profiles of the three main modes is presented during their transformation along the dispersion curve associated to the increase of $E_0$. The field profiles of the S0 and AN0 modes do not change qualitatively. On the contrary, the field profile of the AS1 mode undergoes a qualitative transformation. For low $E_0$ values, this mode is highly asymmetric and strongly localized on the right core interface $x=d$. With the increase of~$E_0$ the asymmetric profile becomes more symmetric, and at the point of bifurcation it perfectly overlaps with the symmetric mode [see Fig.~\ref{fig:fields-Hy-sym0-anti0-asym1}(c)]. For $E_0$ values above the bifurcation point, the mode becomes asymmetric and it tends to localize on the left interface.

In Fig.~\ref{fig:fields-Hy-symI-asym2} a similar transformation is shown for the SI and AS2 modes. Here, with the increase of $E_0$, the peak amplitude of the soliton $H_{\textrm{peak}} = H_y(x=d/2)$ first decreases (it is the lowest at the bifurcation) and then increases, while side lobe peak amplitude  $H_{\textrm{lobe}}$ (located at $x=0$ and $x=d$) of the symmetric mode increases monotonously with $E_0$. The ratio  $H_{\textrm{peak}}/H_{\textrm{lobe}}$ first decreases (up to the bifurcation) and then increases. In case of the asymmetric mode AS2, with the increase of $E_0$ the soliton peak shifts from left to right. At the same  time the amplitude of the left (right) side lobe increases (decreases).
More results on the mode transformation, including the results obtained using the JEM are presented in Ref.~\cite{Walasik-thesis14}.

\subsection{Single interface limit}
\label{sec:results-single-interface-limit}
In Sec.~\ref{sec:ver} in Ref.~\cite{Walasik14b}, describing the theoretical derivation of the models for NSWs, we mentioned that in the limiting case, where the integration constants $c_0$ in Eqs.~(\ref{eqn:first_int}) and (\ref{eqn:c0}) or $C_0$ in Eqs.~(\ref{eqn:IM0}) and (\ref{eqn:IM_disp0}) (all equation numbers correspond to Ref.~\cite{Walasik14b}) are equal to zero, we recover the case of a single interface between a metal and a nonlinear dielectric. Looking at the field profiles of highly asymmetric modes AS1 [see Figs.~\ref{fig:fields-Hy-sym0-anti0-asym1}(a), (e)], we see that these modes are mostly localized at one interface only. Therefore, they can be well approximated by a solution of the single-interface problem.

\begin{figure}[!t]
	\centering
	\includegraphics[width=1.0\columnwidth,clip=true,trim= 0 0 0 0]{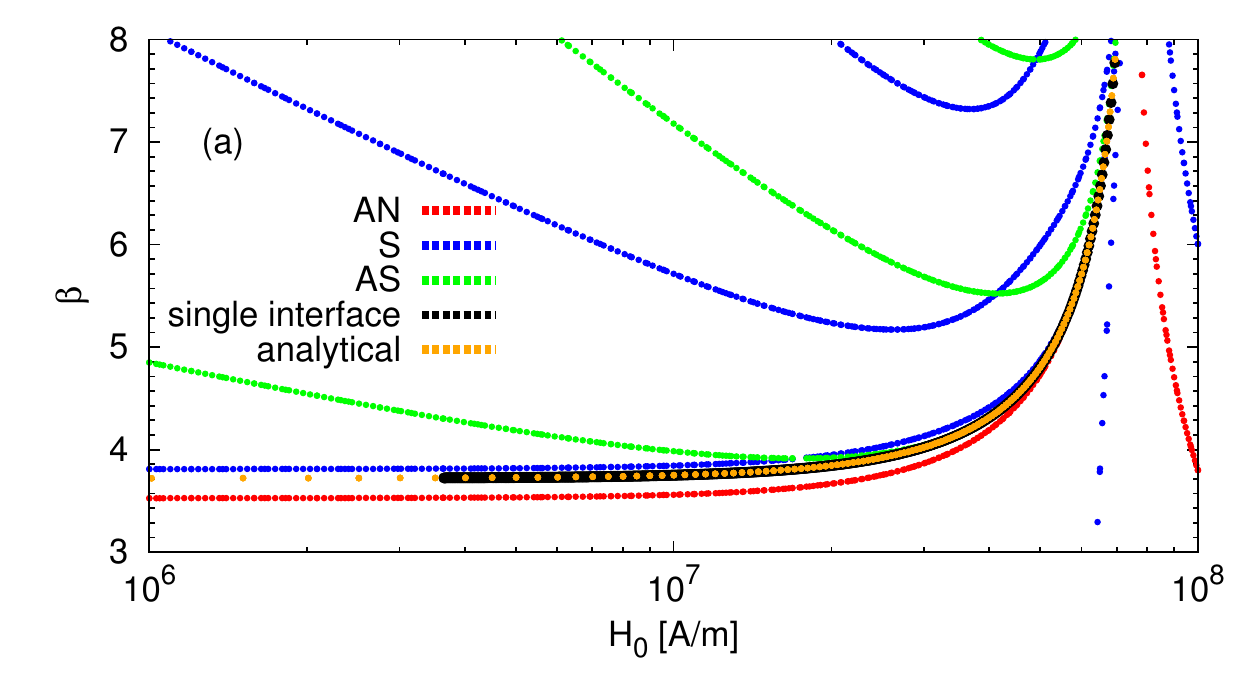}
	\includegraphics[width=1.0\columnwidth,clip=true,trim= 0 0 0 0]{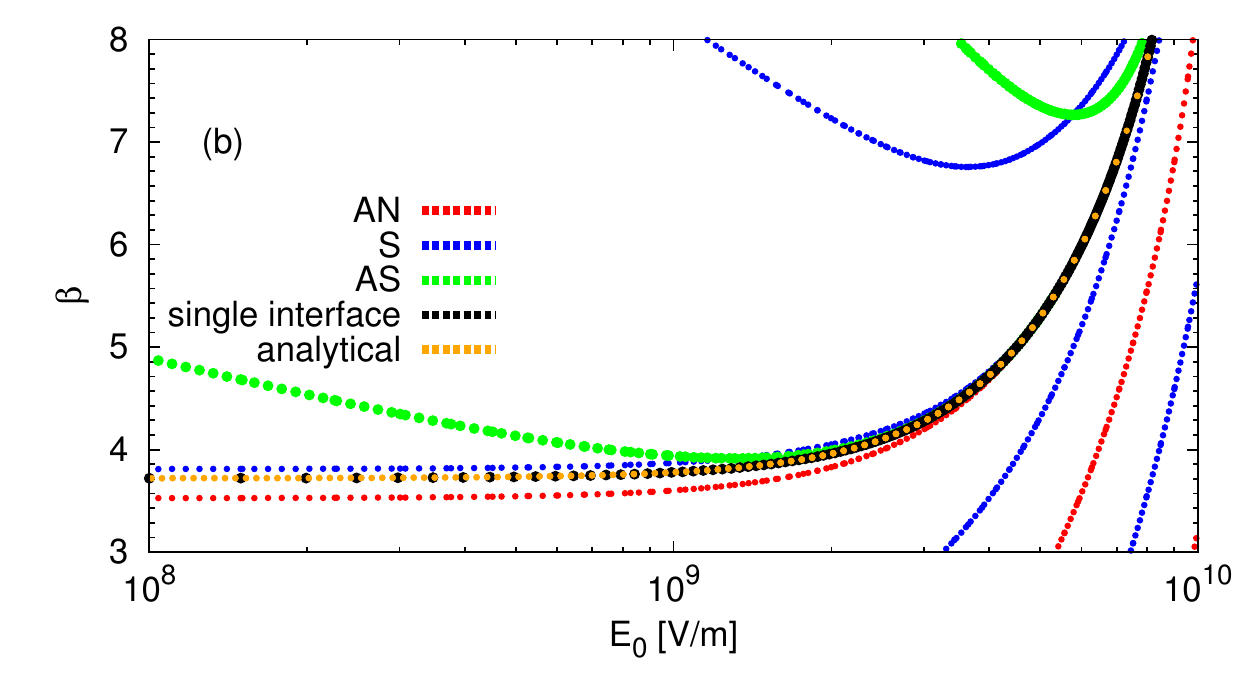}
	\caption{Dispersion diagrams (a) $\beta(H_0)$ obtained using JEM and (b) $\beta(E_0)$ obtained using the IM for the  symmetric NSW. Dispersion curves presenting single-interface approximation obtained using models derived in Ref.~\cite{Walasik14} are shown in black. Additionally, the curves corresponding to the analytical expression for the single-interface dispersion relation \{Eqs.~(\ref{eqn:beta_approx}) and (\ref{eqn:IM_single-anal}) in Ref.~\cite{Walasik14b}\} are shown in yellow.}
	\label{fig:disp-single}
\end{figure}

In Fig.~\ref{fig:disp-single}, we present the dispersion curves for the NSW obtained using the JEM [$\beta(H_0)$]   \{see Eq.~(\ref{eqn:met_fields_a}) in Ref.~\cite{Walasik14b}\} and the IM [$\beta(E_0)$] (compare with Fig.~\ref{fig:disp-E0}). Additionally to the antisymmetric (red), symmetric (blue), and asymmetric (green) dispersion curves, black dispersion curves obtained using single-interface models are presented.
In the case of the JEM, the single-interface approximation was obtained using the 'field based model' for configurations with semi-infinite nonlinear medium  described in Ref.~\cite{Walasik14}. This model was used for a single interface between a metal and a nonlinear dielectric with the same parameters as our NSW.
In case of the IM, the corresponding single-interface approximation was obtained using the 'exact model' for configurations with a semi-infinite nonlinear medium described in Ref.~\cite{Walasik14}. 

In Fig.~\ref{fig:disp-single}, we see that both for the JEM and for the IM, the single-interface dispersion curve always lays between the antisymmetric AN0 curve and the symmetric S0 curve. For high values of $E_0$, the asymmetric AS1 curve becomes very close to the black curve, but remains slightly above it.
The fact that the black curves overlap with the green AS1 curves confirms that the highly asymmetric AS1 modes (for high effective index $\beta$) are well approximated using the single-interface approach.

Now, instead of using the corresponding models for configurations with a semi-infinite nonlinear medium, we will use the formulas found in Sec.~\ref{sec:ver} in Ref.~\cite{Walasik14b} that give us the analytical expressions for the dispersion relations for the single-interface problem. 
In the case of the JEM, the analytical formula for the dispersion relation of a single metal/nonlinear dielectric interface problem is given by Eq.~(\ref{eqn:beta_approx}) in Ref.~\cite{Walasik14b}. In this equation, as in the entire formulation of the JEM, the primary parameter is the magnetic field amplitude at the interface $H_0$. Therefore, we are able to show the dependency described by Eq.~(\ref{eqn:beta_approx}) only in the coordinates where the effective index is presented as a function of the magnetic field amplitude at the interface $H_0$ [see yellow line in Fig.~\ref{fig:disp-single}(a)].  We observe that the dispersion relations calculated using the field based model (balck curve) and the yellow curve described by Eq.~(\ref{eqn:beta_approx}) overlap perfectly. The single-interface dispersion curve, which corresponds to the limiting case $c_0 = 0$ divides the dispersion plot $\beta(H_0)$ into the regions corresponding to the node-less family and the family with nodes as predicted in Section~\ref{sec:JEM-theory}  in Ref.~\cite{Walasik14b}
Above the $c_0=0$ curve (for negative values of the integration constant $c_0$), only node-less solutions exist. Below the $c_0=0$ curve (for $c_0 >0$), only solutions with nodes exist.

In case of the IM, the analytical formula for the dispersion relation for the single-interface problem is given by Eq.~(\ref{eqn:IM_single-anal}) in Ref.~\cite{Walasik14b}. The effective index of the mode expressed as a function of the material parameters of the structure and the total electric field intensity at the interface $E_0$. The curve described by Eq.~(\ref{eqn:IM_single-anal}) is plotted in yellow in Fig.~\ref{fig:disp-single}(b) and it overlaps well with the black curve obtained using the exact model.

In case of the IM, the numerical results also show that the dispersion curves are divided in two families: with nodes and node-less. The regions of the dispersion diagram corresponding to these two families are separated by the curve described by the equation $C_0=0$ (black and yellow curves for the single-interface problem). In the frame of the IM, we could not prove this property analytically because the field plots in the IM are calculated numerically.

In Fig.~\ref{fig:disp-single}, in the region of high effective indices, the dispersion curves of the AS1 mode overlap with the curves obtained using single-interface approximations. This confirms our hypothesis that highly asymmetric modes AS1 can be approximated by solutions obtained using the corresponding single-interface models.

\subsection{Comparison with linear states}
\label{sec:comp-lin}

In Sec.~\ref{sec:res-sym-disp}, while discussing the field profiles of the modes belonging to the family with nodes, we noticed that they resemble higher-order modes of linear slot waveguides with parameters similar to the NSW studied here. In this section, we will explain the origin of the similarities between these nonlinear and linear modes.

\begin{figure}[!b]
	\includegraphics[width=1\columnwidth,clip=true,trim= 0 0 0 0]{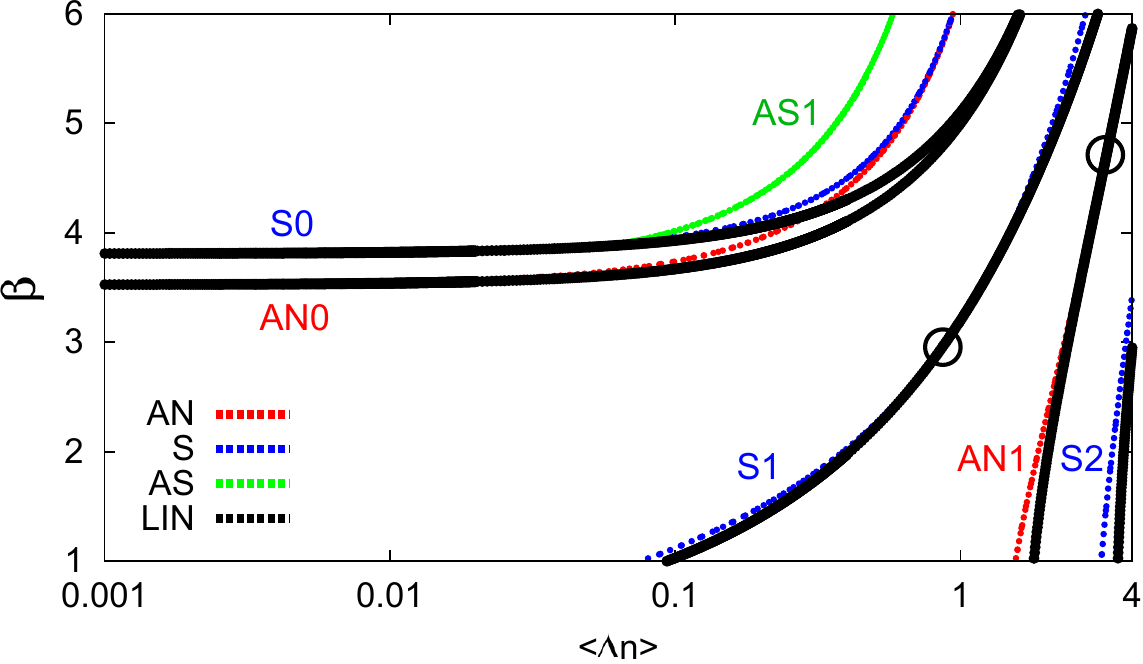}
	\caption{A comparison of the nonlinear (red, blue and green curves) and the linear dispersion plots (black curves) of the symmetric slot waveguides. In case of the linear waveguide $\langle \Delta n \rangle$ is equivalent to $\Delta n_{\textrm{lin}}$. Points correspond to the modes presented in Fig.~\ref{fig:equiv-flat-profiles}.}
	\label{fig:disp-flat-comp}
\end{figure}

In Fig.~\ref{fig:disp-flat-comp}, we present the nonlinear dispersion diagram obtained using the IM for our NSW. In this plot, the effective index of the mode $\beta$ is presented as a function of the averaged nonlinear index modification in the waveguide core $\langle \Delta n \rangle$: 
\begin{equation}
\langle \Delta n \rangle = \frac{1}{d} \int_0^d \Delta n \mathrm{d}x = \frac{1}{d} \int_0^d n_2^{(2)} I \mathrm{d}x,
\label{eq:definition-delta-n}
\end{equation}
where the nonlinear parameter $n_2^{(2)} = \alpha_2/\epsilon_0 c \epsilon_{l,2}$


In addition to this plot we present also a dispersion relation (black curves in Fig.~\ref{fig:disp-flat-comp}) of a linear slot waveguide with a homogeneous and linear core and the following parameters: $\epsilon_1 = \epsilon_3 = -90$, $n = n_0 + \Delta n_{\textrm{lin}} = 3.46+ \Delta n_{\textrm{lin}}$ and $d = 400$~nm. The parameters $\epsilon_1$, $\epsilon_3$ and $n_0 = \sqrt{\epsilon_{l,2}}$ are identical to these in the case of the nonlinear waveguide studied here.

We notice that this linear dispersion diagram is similar to the dispersion plot of the NSW. For the core with index $n=n_0$ only two modes are present and they are the linear counterparts of the modes S0 and AN0. With the increase of the core index $n$, the effective index of these modes increases and they become closer to each other. At $\Delta n_{\textrm{lin}} \approx 0.1$, a higher-order linear mode appears that is a counterpart of the S1 mode. For  $\Delta n_{\textrm{lin}} \approx 2$ and $\Delta n_{\textrm{lin}} \approx 3.5$, another two higher-order modes appear. They are the linear counterparts of the AN1 and S2 modes, respectively. The effective index of these modes increases rapidly with the increase of $\Delta n_{\textrm{lin}}$. The only modes not present in the linear dispersion curves are the asymmetric modes AS1, AS2, \dots and the symmetric node-less modes SI, SII, etc. The asymmetric modes can not be observed in the linear case because nothing breaks the symmetry in the symmetric linear slot waveguide. The node-less symmetric modes are not supported by the linear slot waveguide because they have purely nonlinear solitonic character [see Figs.~\ref{fig:fields-Hy-symI-asym2} and \ref{fig:IM-field-examples}(e)].

The dispersion curves of the nonlinear modes AN0 and S0 overlap with the corresponding linear dispersion curves only for small $\langle \Delta n \rangle$ values. The nonlinear modes increase their effective indices $\beta$ faster than the linear modes. In case of higher-order modes S1, AN1 and S2, the dispersion curves of the linear modes lay below the corresponding nonlinear modes. There is only one common point per mode for these curves (indicated by an open circle in Fig.~\ref{fig:disp-flat-comp}) and it turns out that at this point, the index distribution induced by the nonlinear mode in the nonlinear core is flat (data not shown). 

\begin{figure}[!t]
	\centerline{\includegraphics[width=0.50\columnwidth,clip=true,trim= 0 0 0 0]{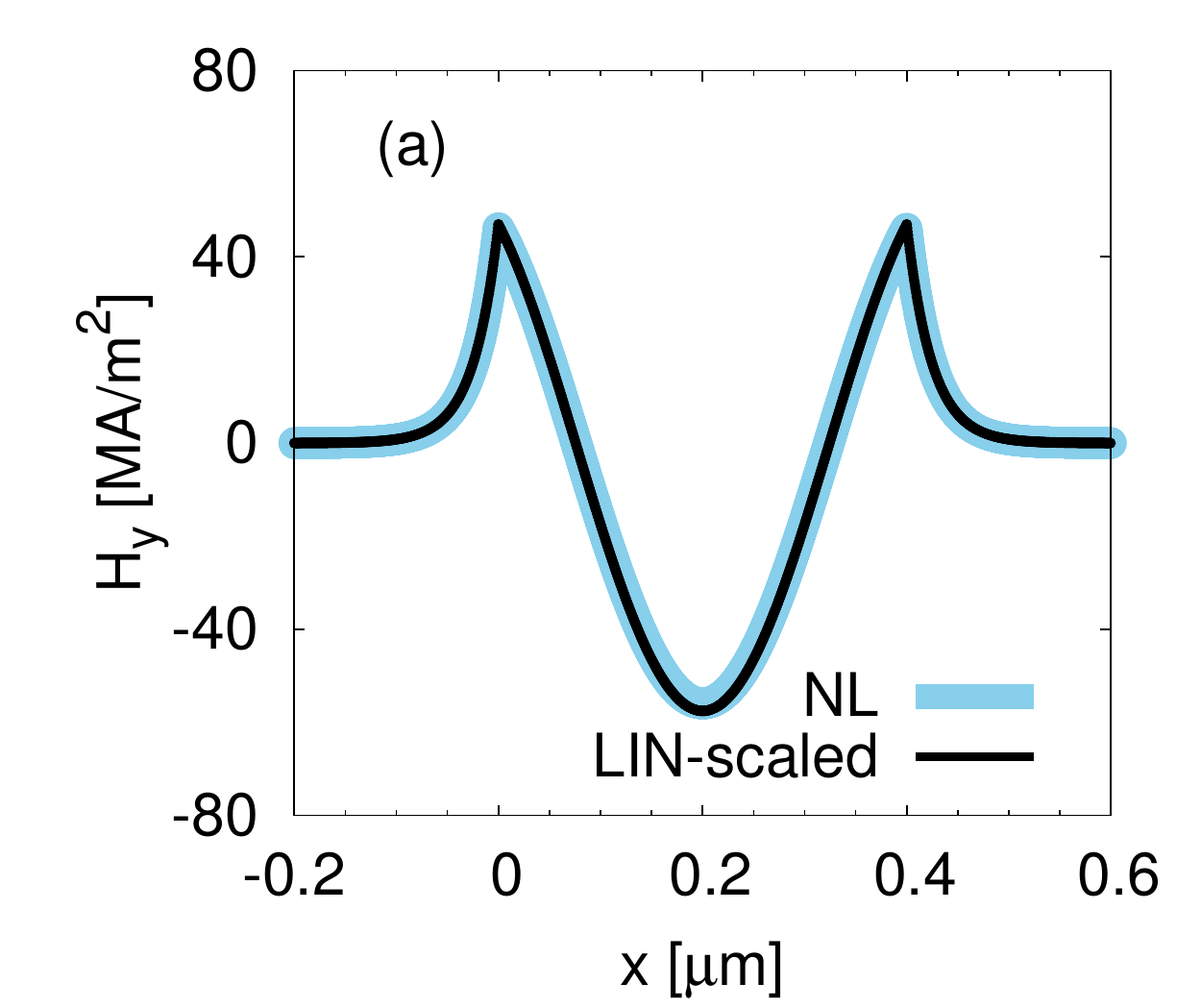}
		\includegraphics[width=0.50\columnwidth,clip=true,trim= 0 0 0 0]{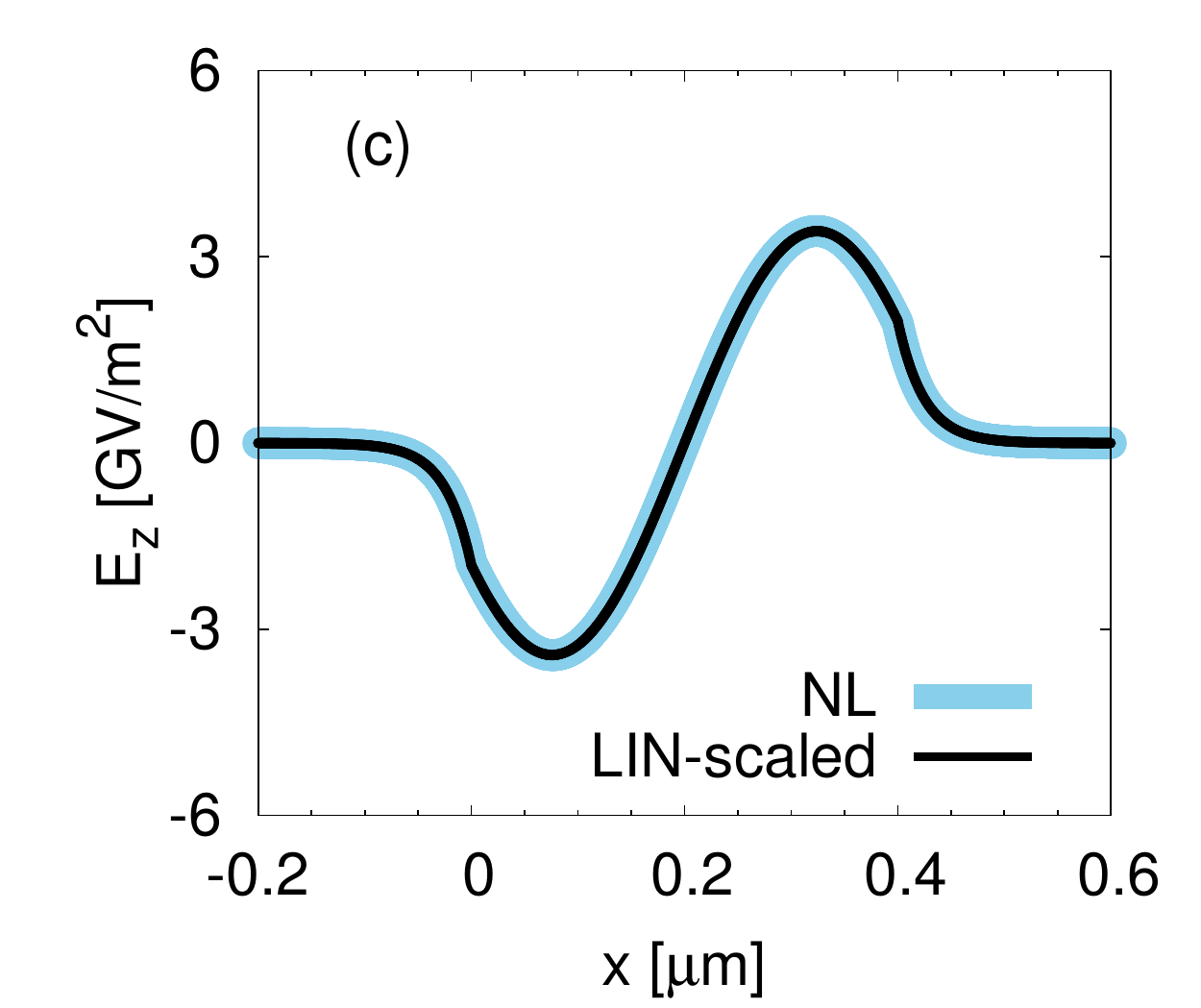}}
	\centerline{\includegraphics[width=0.50\columnwidth,clip=true,trim= 0 0 0 0]{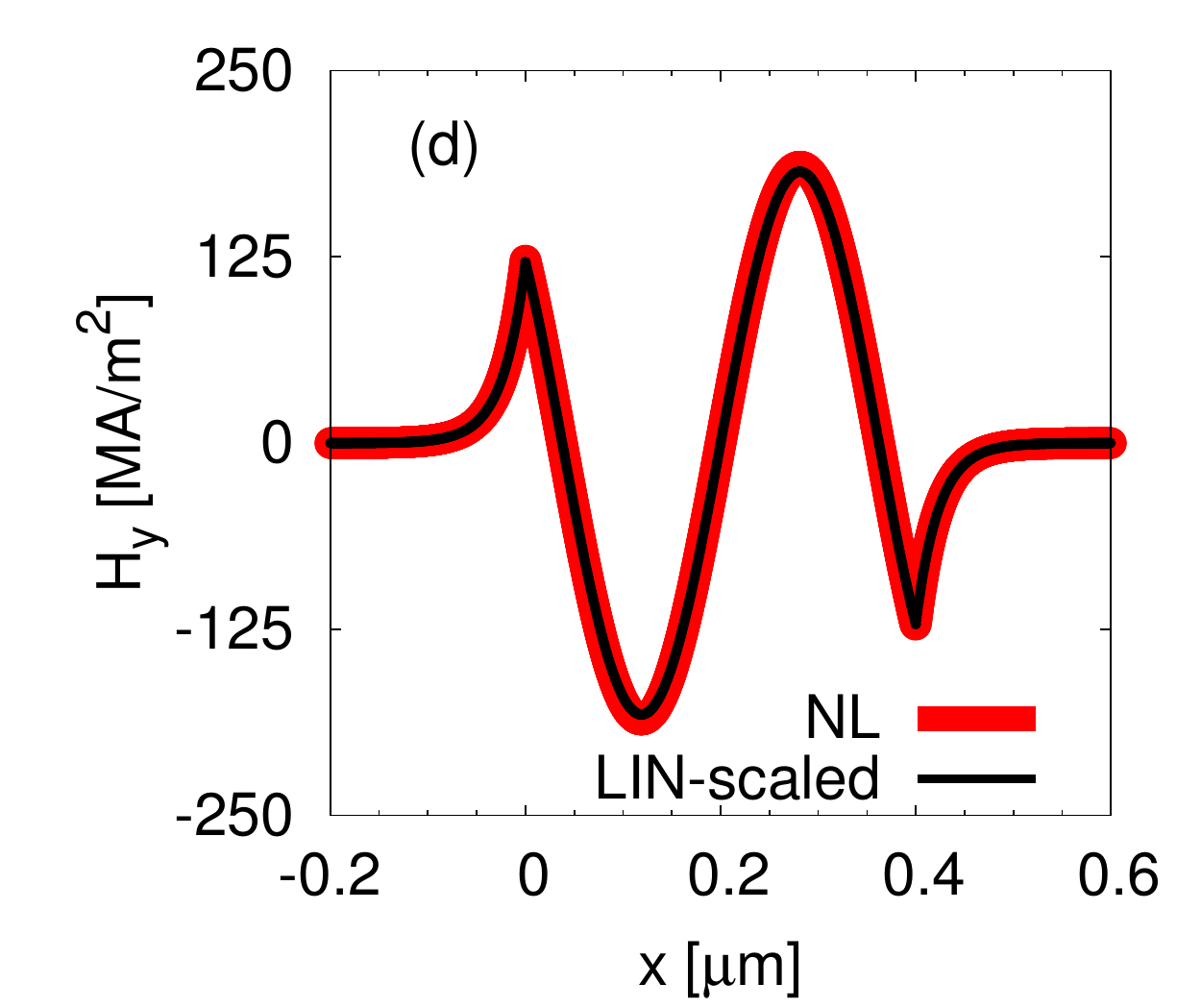}
		\includegraphics[width=0.50\columnwidth,clip=true,trim= 0 0 0 0]{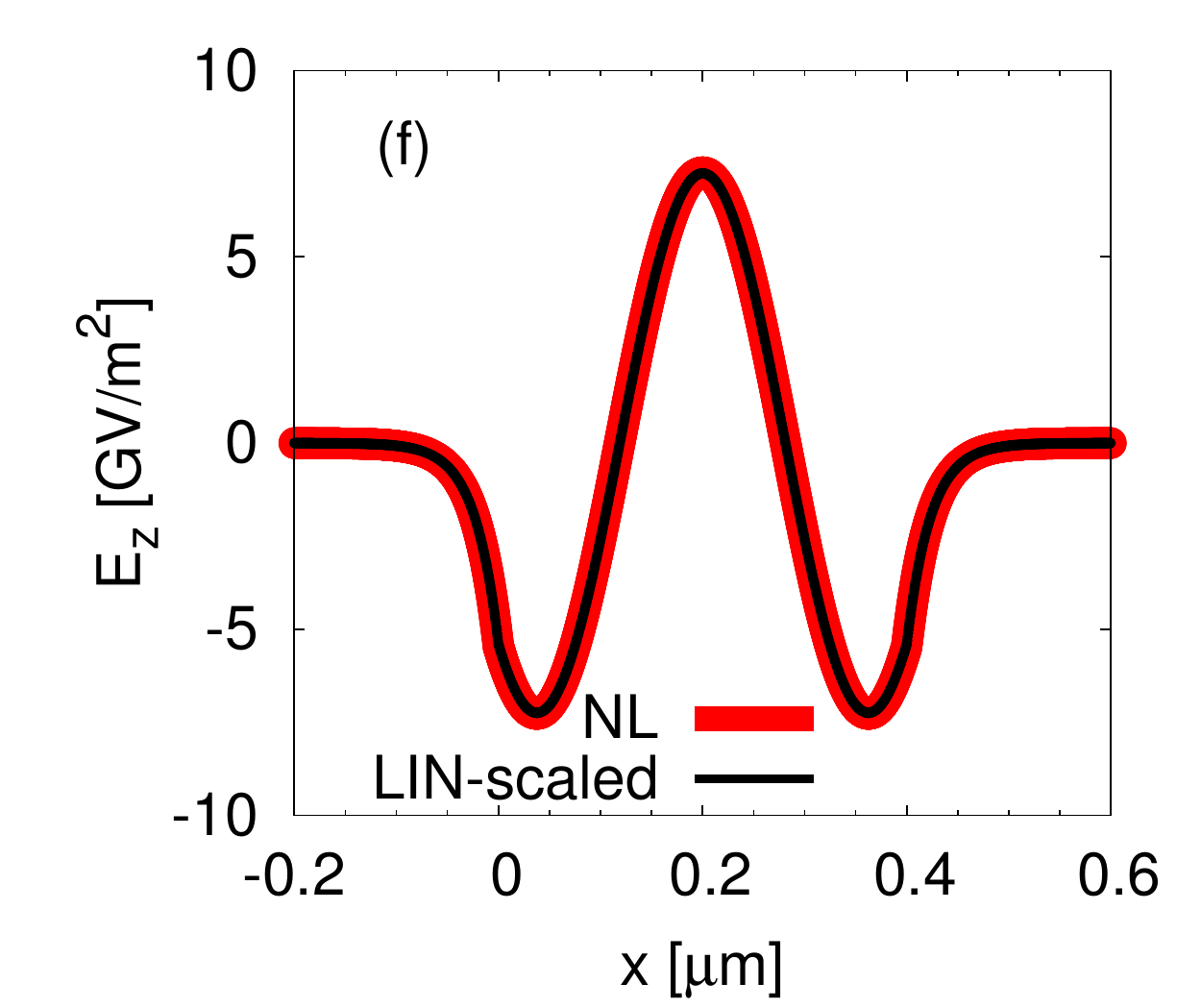}}
	\caption{A comparison of (a), (c) $H_y(x)$ and (b), (d) $E_z(x)$ for the nonlinear modes S1 (blue curve) and AN1 (red curve) and the normalized profiles of their linear counterparts (black curves) at the common points of the black (for linear slot waveguide) and blue or red (for NSW) dispersion curves indicated by open circles in Fig.~\ref{fig:disp-flat-comp}.}
	\label{fig:equiv-flat-profiles}
\end{figure}

Figure~\ref{fig:equiv-flat-profiles} present the comparison of the field profiles $H_y(x)$ and $E_z(x)$ for nonlinear S1 and AN1 modes and their linear counterparts, at the points where the index distribution induced by the nonlinear mode in the nonlinear core is flat. We observe that the nonlinear profiles overlap perfectly with the profiles of the linear modes normalized to the same amplitude as the nonlinear modes.

The results presented here prove that the modes with nodes found in the NSW are close to the modes of the linear slot waveguide with similar opto-geometric parameters. We explain the similarities between these nonlinear and linear modes using the self-coherent definition of nonlinear modes. This definition was introduced by Townes and co-workers in Ref.~\cite{Chiao64} and was used later in other works (\textit{e.g.}, Ref.~\cite{Mitchell93}). It defines a nonlinear mode as a linear mode of a linear (graded refractive index) waveguide that is induced by the light distribution of this mode. According to this definition, there is no difference between the nonlinear modes of the NSW for which the nonlinear index modification has a flat distribution and the linear modes of the waveguide with higher, uniformly distributed refractive index of the linear core.

\subsection{Permittivity contrast}

In Ref.~\cite{Walasik14a}, we have studied the influence of the width of the NSW core on the nonlinear dispersion for this this structure. Here we will discuss the influence of the permittivity contrast between the dielectric core and the metal cladding on the nonlinear dispersion diagrams of symmetric NSW.

\begin{figure}[!t]
	\centerline{\includegraphics[width=0.49\columnwidth,clip=true,trim=15 0 0 0]{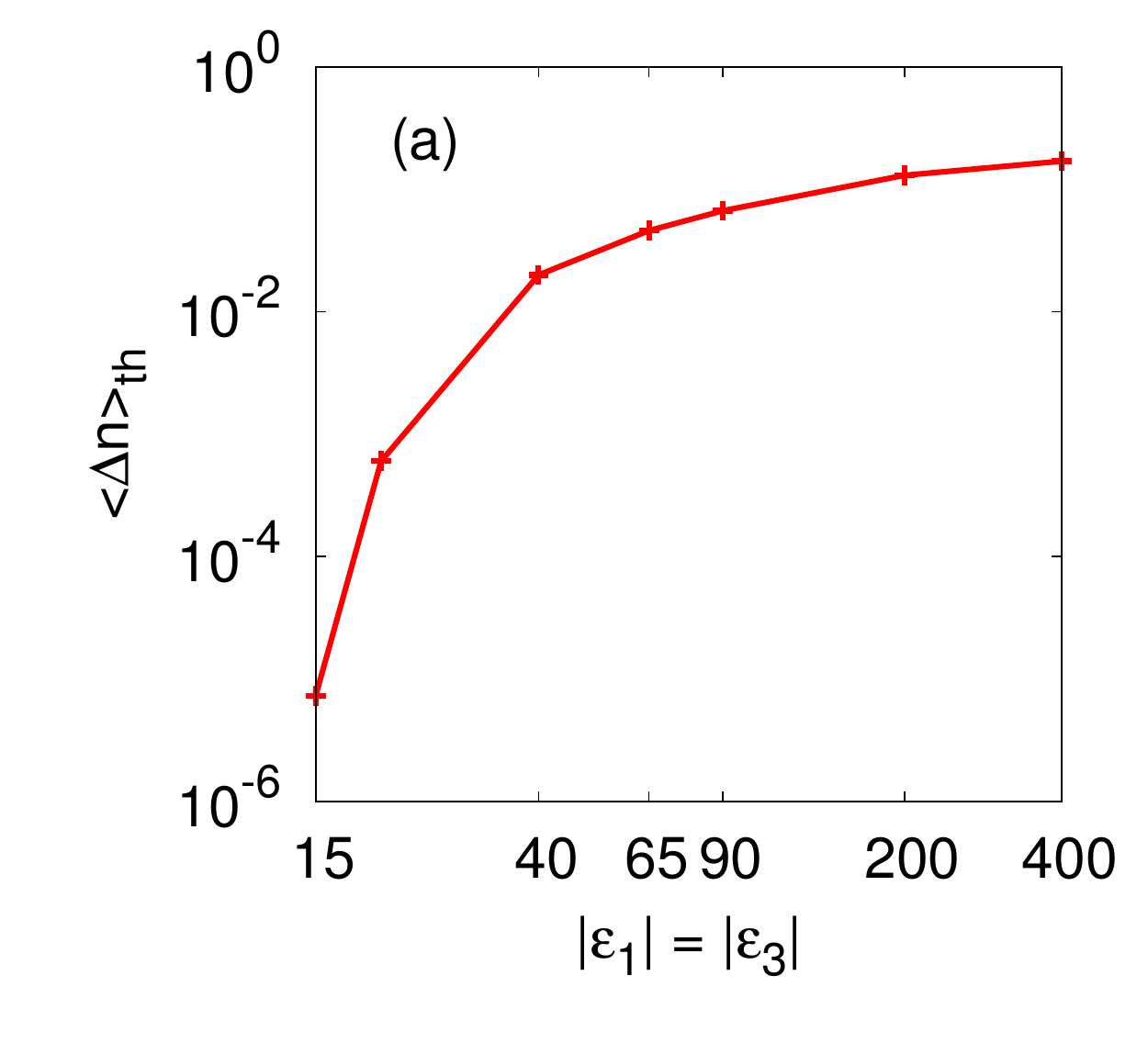}
		\includegraphics[width=0.49\columnwidth,clip=true,trim=15 0 0 0]{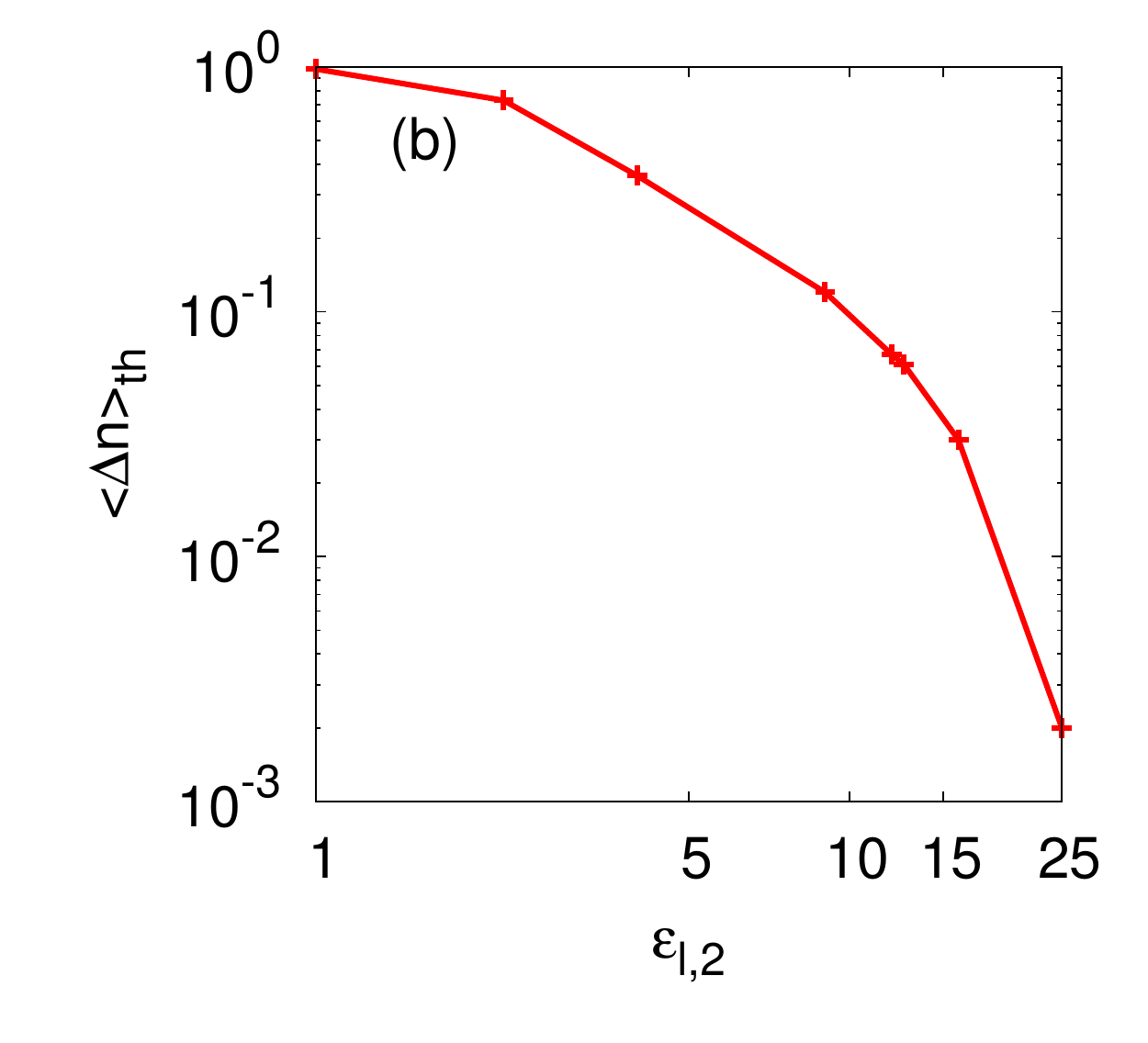}}
	\caption{Average nonlinear index change at the appearance of the asymmetric AS1 modes $\langle\Delta n\rangle_{\textrm{th}}$ as a function of the absolute value of (a) the metal cladding permittivity of the symmetric waveguide $|\epsilon_1| = |\epsilon_3|$ and (b) the linear part of the nonlinear core permittivity $\epsilon_{l,2}$ All the other parameters of the NSW are identical to these used in Sec.~\ref{sec:res-sym-disp}.}
	\label{fig:index-contrast-metal}
\end{figure}

First, we will discuss the influence of the metal cladding permittivity on the nonlinear dispersion diagrams of NSW. We have studied the dispersion plots for the NSWs with identical parameters as these used in Sec.~\ref{sec:res-sym-disp} but with different values of the metal cladding permittivity. We observe that the cladding with higher permittivity  (lower in absolute value) allows us to reduce the $\langle\Delta n\rangle$ threshold values where the bifurcation of the AS1 mode occurs. For metals with permittivity equal to~$-40$, the bifurcation occurs at $\langle\Delta n\rangle \approx 0.02$, which is 4 times lower than in the case of $\epsilon_1 = \epsilon_3 = -90$. For metals with permittivity equal to $-20$, the bifurcation occurs at $\langle\Delta n\rangle \approx 8 \cdot 10^{-4}$, which corresponds to the reduction of the bifurcation threshold by two orders of magnitude with respect to the configuration with $\epsilon_1 = \epsilon_3 = -90$. The dependency of the AS1 mode bifurcation threshold $\langle \Delta n \rangle_{\textrm{th}}$ on the metal cladding permittivity is illustrated in Fig.~\ref{fig:index-contrast-metal}(a). Looking at this plot, we conclude that with the increase of the metal cladding permittivity (decrease of its absolute value) the bifurcation threshold of the AS1 mode decreases. This decrease is slow in the range of high index contrast between the metal and the nonlinear dielectric permittivity. Although, for $|\epsilon_1| = |\epsilon_3|$ close to $\epsilon_{l,2}$ the decrease of the bifurcation threshold is more rapid. Changing the metal permittivity from $-20$ to $-15$ allows us to decrease the bifurcation threshold by almost two order of magnitudes. For the metal cladding permittivity $\epsilon_1 = \epsilon_3 =  -15$ the bifurcation threshold is at the level of $\langle\Delta n\rangle \approx 10^{-5}$. This is four orders of magnitude lower than for the $\epsilon_1 = \epsilon_3$ from the range $[-400,-90]$, for which the bifurcation occurs at $\langle\Delta n\rangle \approx 0.1$. 

Next, we will study the influence of the change of the core permittivity  on the dispersion diagram of the symmetric NSW.
We analyzed the plots of the dispersion curves for the NSWs with different linear parts of the core permittivity $\epsilon_{l,2}$. All the other parameters are identical to these used in Sec.~\ref{sec:res-sym-disp}. The behavior of the bifurcation threshold expressed as the averaged nonlinear index modification $\langle \Delta n \rangle$ is presented in Fig.~\ref{fig:index-contrast-metal}(b). The increase of the linear part of the core permittivity $\epsilon_{l,2}$ is accompanied by a monotonous decrease of the bifurcation threshold. From Fig.~\ref{fig:index-contrast-metal}(b) we notice that the increase of $\epsilon_{l,2}$ from 1 to 25 results in the decrease of the bifurcation threshold by approximately three orders of magnitude.

It is interesting to remind that, in the case of changing the permittivity contrast by varying the metal cladding permittivity [see Fig.~\ref{fig:index-contrast-metal}(a)], we observed a decrease of the bifurcation threshold for the AS1 mode with the decrease of the permittivity contrast between the cladding and the core permittivity. On the contrary, decreasing the permittivity contrast by changing the core permittivty, leads to the increase of the bifurcation threshold [see Fig.~\ref{fig:index-contrast-metal}(b)].

This phenomenon can be explained using the field profiles of the symmetric mode for different values of core and metal permittivities. We observe that increasing the permittivity of the core or increasing the permittivity of the metal (decreasing its absolute value) leads to symmetric modes that are more localized on the waveguide interfaces and look closer to separate plasmon on both metal/dielectric interfaces. Because the overlap and therefore the interaction between the two plasmons is weaker, it is easier to break the symmetry of the mode. This explains the decrease of the bifurcation threshold. 
We conclude that changing the permittivity contrast by varying the linear part of the nonlinear core permittivity, has opposite effect than changing the permittivity contrast by varying the metal cladding permittivity.

\section{Results for asymmetric structures}
\label{sec:res-asym}

In Sec.~\ref{sec:res-sym}, we have comprehensively discussed dispersion diagrams and mode profiles in symmetric NSW structures. In this section, we will discuss the influence of the NSW asymmetry on the dispersion curves. The asymmetry is introduced by sandwiching the nonlinear core by metals with different values of the permittivity on both sides. Asymmetric NSW structures have not been studied before in literature. Here we present the analysis of these structures for the first time.

\subsection{Dispersion relations}

\begin{figure}[!h]
	\centerline{\includegraphics[width=1\columnwidth,clip=true,trim=0 10 0 70]{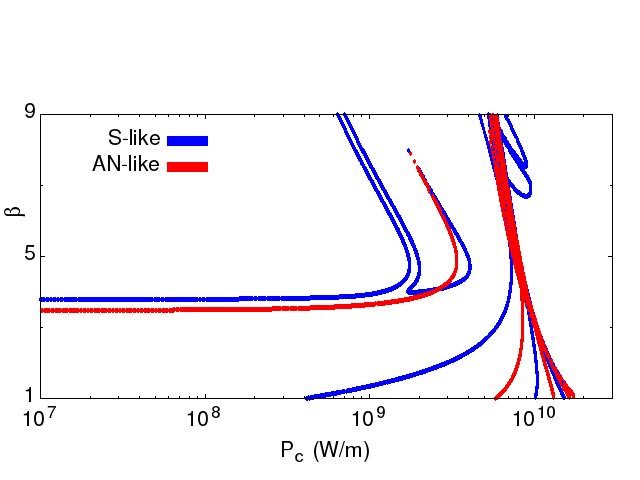}}
	\caption{Dispersion diagram obtained using the IM for the asymmetric structure with $\epsilon_1 = -110$ and $\epsilon_3 = -90$ (for the scheme of the structure see Fig.~\ref{fig:geometry-slot}). Blue curves correspond to the modes for which $\sign[E_{x,0}] = \sign[E_{x,d}]$ and red curves correspond to the modes for which $\sign[E_{x,0}] = - \sign[E_{x,d}]$ \{see Eq.~(\ref{eqn:total-elect}) in Ref.~\cite{Walasik14b} for the notations\}. Compare this dispersion diagram for the asymmetric structure with the dispersion diagram for the symmetric structure presented in Fig.~\ref{fig:disp-Pc}(b).} 
	\label{fig:asym-disp-dn-110}
\end{figure}

Figure~\ref{fig:asym-disp-dn-110} presents the nonlinear dispersion diagram obtained using the IM for the structure with the following parameters: core permittivity $\epsilon_{l,2} = 3.46^2$, the second-order nonlinear refractive index $n_2^{(2)} = 2\cdot10^{-17}$~m$^2$/W, core with $d =400$~nm, metal permittivities $\epsilon_1 = -110$, $\epsilon_3 = -90$ at a free-space wavelength $\lambda = 1.55$~$\mu$m. These parameters are identical to those for the structure studied in Sec.~\ref{sec:res-sym-disp} except for the metal permittivities. Here the permittivity of the left metal layer is decreased to $-110$ making the structure asymmetric.

\begin{figure}[!ht]
	\includegraphics[width=1\columnwidth,clip=true,trim= 0 10 0 70]{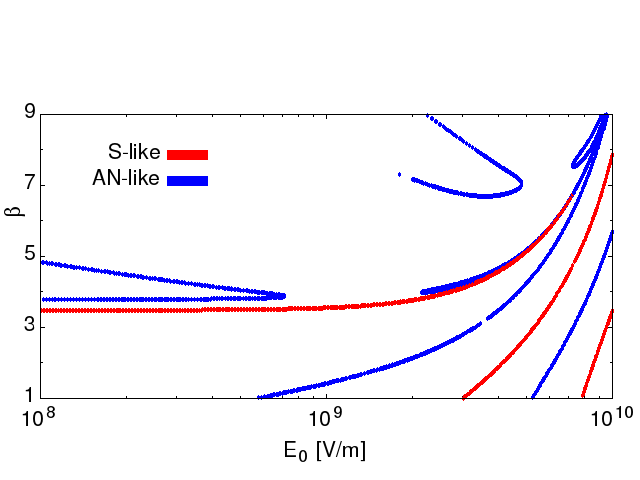}
	\caption{Dispersion curves  $\beta(E_0)$ for the asymmetric structure with $\epsilon_1 = -110$ and $\epsilon_3 = -90$. Compare this dispersion diagram for the asymmetric structure with the dispersion diagram for the symmetric structure presented in Fig.~\ref{fig:disp-E0}.}
	\label{fig:IM-disp-E0-mode-class-asym}
\end{figure}

In the asymmetric structure only asymmetric modes are present. However, in the dispersion diagram shown in Fig.~\ref{fig:asym-disp-dn-110}, we divide the modes in two groups: modes that resemble the antisymmetric modes of the symmetric structure for which $\sign[E_{x,0}] = - \sign[E_{x,d}]$ (red curves labeled AN-like) and modes that resemble the symmetric or asymmetric modes of the symmetric structure for which $\sign[E_{x,0}] = \sign[E_{x,d}]$ (blue curves labeled S-like)  \{see Eq.~(\ref{eqn:total-elect}) in Ref.~\cite{Walasik14b} for the notations of the electric field components\}.

We compare the nonlinear dispersion curves for the asymmetric structure presented in Fig.~\ref{fig:asym-disp-dn-110} with the dispersion curves obtained for the symmetric structure shown in Fig.~\ref{fig:disp-Pc}. We notice that the dispersion curves for the symmetric and antisymmetric modes from the family with nodes did not change much. The number of modes and the character of their dispersion curves is conserved. 
The main difference between the dispersion curves of the asymmetric and symmetric structures can be observed for the symmetric and asymmetric modes of the node-less family. The asymmetry of the structure lifts the double degeneracy of the asymmetric branch AS1 (see green curve in Fig.~\ref{fig:disp-Pc}). This branch splits into two branches (see Fig.~\ref{fig:asym-disp-dn-110}). One of them (the branch with lower effective indices $\beta$) is a continuation of the symmetric-like fundamental mode (blue curve) that starts for small power density $P_c$ levels. The second branch lays along the first one but has slightly higher power levels (branch with higher $\beta$ values). The degeneracy of the higher-order asymmetric modes is also lifted by the asymmetry of the structure. These branches also split into two separate branches, similar to the case of the AS1 mode. It is difficult to observe this effect in Fig.~\ref{fig:asym-disp-dn-110}, where the power density in the core is used as abscissa (even enlarging the region of interest), because the two dispersion curves into which the dispersion curve of the higher-order asymmetric mode splits lay very close to each other. The degeneracy lift of the AS2 mode can be however observed from the dispersion curve $\beta(E_0)$ presented in Fig.~\ref{fig:IM-disp-E0-mode-class-asym}, where the effective index is shown as a function of the field intensity at the left core interface. In these coordinates the separation of the SI and AS2 curves reflects the degeneracy lift of the AS2 mode.

\subsection{Permittivity contrast study}

\begin{figure}[!h]
	\centerline{\includegraphics[width=1\columnwidth,clip=true,trim= 0 10 5 70]{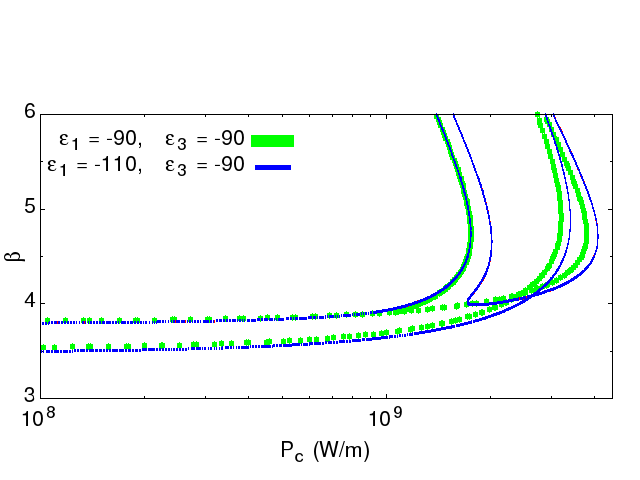}}
	\caption{Dispersion curves of the asymmetric NSW with $\epsilon_1 = -110, \epsilon_3 = -90$ (blue curves) and the symmetric structure $\epsilon_1 =  \epsilon_3 = -90$ (green curves).}
	\label{fig:asym-disp-comp-110-90}
\end{figure}

To finish our discussion of the asymmetric NSW properties, we directly compare the dispersion diagrams $\beta(P_c)$ of the symmetric structure with these of the asymmetric structures. In Fig.~\ref{fig:asym-disp-comp-110-90}, the dispersion plot of the symmetric structure ($\epsilon_1 = \epsilon_3 = -90$, see Fig.~\ref{fig:disp-Pc}) is compared with the dispersion plot for the asymmetric structure ($\epsilon_1 = -110, \epsilon_3 = -90$, see Fig.~\ref{fig:asym-disp-dn-110}). Only a vicinity of the bifurcation point of the AS1 mode is presented. 
We observe that, for low $P_c$ values, the dispersion curves of the two low-power modes are slightly modified due to the waveguide asymmetry. For higher values of $P_c$, the dispersion curve of the  fundamental mode (upper blue curve) exactly overlaps with the dispersion curve of the asymmetric mode of the symmetric structure. This is a consequence of the fact that the field profiles corresponding to this upper blue curve  are strongly localized on the interface with the metal with higher value of the permittivity. These profiles resemble the profiles of the highly asymmetric modes of the symmetric structure [see Fig.~\ref{fig:fields-Hy-sym0-anti0-asym1}(a)]. Therefore, we are not surprised that these two dispersion curves overlap. The second curve that results from the degeneracy lift of the asymmetric mode lays above (in terms of $P_c$) the dispersion curve of the asymmetric mode AS1  (green curve).

\begin{figure}[!h]
	\centerline{\includegraphics[width=1\columnwidth,clip=true,trim= 0 10 5 70]{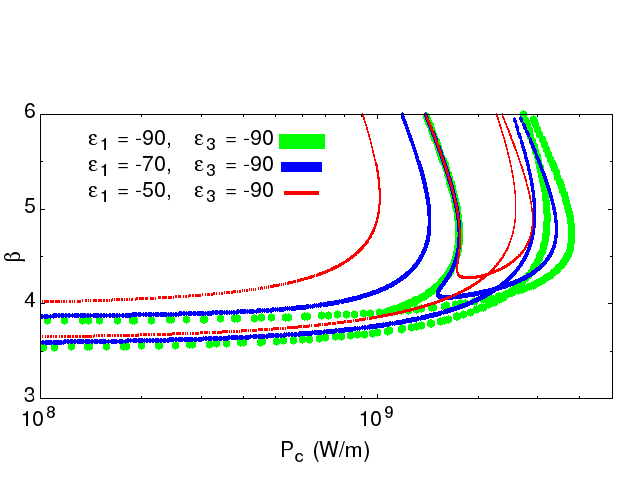}}
	\caption{Dispersion curves of the asymmetric NSWs with $\epsilon_1 = -70$ and $\epsilon_3 = -90$ (blue curves), $\epsilon_1 = -50$ and $\epsilon_3 = -90$ (red curves), and the symmetric structure $\epsilon_1 = \epsilon_3 = -90$ (green curves).}
	\label{fig:asym-disp-comp-50-70-90}
\end{figure}

In Fig.~\ref{fig:asym-disp-comp-50-70-90}, we present a comparison of the dispersion curves of the symmetric structure ($\epsilon_1 = \epsilon_3 = -90$, see Fig.~\ref{fig:disp-Pc}) and the asymmetric structures, where one of the metal permittivity values is higher than in the case of the symmetric structure. The dispersion curves of the symmetric structure (green curves) are compared with these of the asymmetric structures with $\epsilon_1 = -70$, $\epsilon_3 = -90$ (blue curves), and $\epsilon_1 = -50$, $\epsilon_3 = -90$ (red curves). 

In the case illustrated in Fig.~\ref{fig:asym-disp-comp-50-70-90}, contrary to the one presented in Fig.~\ref{fig:asym-disp-comp-110-90}, it is the higher (in terms of $P_c$) of the two curves that result from the lift of the degeneracy that overlap with the dispersion curve of the asymmetric modes of the symmetric structure. This higher curve corresponds to the modes that are localized on the interface between the core and the metal with permittivity equal to $-90$. For the  structures studied in Fig.~\ref{fig:asym-disp-comp-50-70-90}, $\epsilon = -90$ is the lowest cladding permittivity. For that reason, the dispersion curves corresponding to the mode localized on the interface with metal with lower permittivity, overlap with the dispersion curves of the symmetric structure.

Another effect that can be observed in Fig.~\ref{fig:asym-disp-comp-50-70-90}, is that with the increase of the structure asymmetry $|\epsilon_1 - \epsilon_3|$ the separation of the two curves that appear as a result of the degeneracy lift, increases, as expected. In the limiting case $\epsilon_1 \rightarrow \epsilon_3$, these two curves merge into one doubly degenerate curve.

\section{Stability of the main solutions for symmetric waveguides}
\label{res-stability}
In the previous sections, we have studied the stationary properties of plasmon--soliton waves using two different modal approaches. From both theoretical and practical points of view, the issue of the stability of these waves arises. In several works, the general problem of the stability of nonlinear waves was studied~\cite{Kivshar01,Weilnau02,Kivshar03}. Despite an enormous interest in the properties of nonlinear waves over the last decades, there in no universal condition on their stability~\cite{Mitchell93,Akhmediev97}. In most of the cases, the stability must be studied numerically for each of the cases separately. Stability of nonlinear guided waves in fully dielectric structures was studied numerically in Refs.~\cite{Moloney86,Moloney86a,Mihalache88,Mihalache89,Mihalache94, Mihalache94a,Chiang94,Aceves94,Capobianco95}.

In structures made of metals and nonlinear dielectrics, due to the presence of media with negative permittivity, the problem of stability of plasmon--solitons is difficult to study even numerically. Only in Refs.~\cite{Davoyan09,Milian12} the stability of plasmon--solitons was analyzed for the single nonlinear dielectric/metal interface case, using numerical algorithms (like finite-difference time-domain --- FDTD~\cite{Taflove05,Taflove13}). The propagation of light in plasmonic couplers was studied using Fourier methods based on mode decomposition in linear~\cite{Ctyroky13} and nonlinear~\cite{Petracek13} regimes. In this section, we study the stability of the plasmon--soliton waves in symmetric NSWs using two methodologies: (i) the topological criterion for fundamental modes of nonlinear waveguides derived in Ref.~\cite{Mitchell93} and  (ii) two numerical full-vector methods (using COMSOL~\cite{comsol-web} and nonlinear FDTD implemented in \textsc{meep}~\cite{meep-web,Oskooi10}).

\subsection{Theoretical arguments}
\label{sec:stability-theoretical}

We use here the topological criterion presented in Ref.~\cite{Mitchell93} that is based on the linear stability analysis~\cite{Jones86} and the Vakhitov--Kolokolov criterion~\cite{Vakhitov73}. The stability criterion presented in Ref.~\cite{Mitchell93}, is based only on the topology of the nonlinear dispersion curves and the stability of the modes can be read by analyzing $\beta(I_{\textrm{tot}})$ diagrams in which $I_{\textrm{tot}} \equiv \int^{+\infty}_{-\infty}  I(x) dx$  where $I(x)$ is the intensity density. The validity of this approach was confirmed in multiple settings dealing with purely dielectric structures \cite{Mihalache94, Mihalache94a,Chiang94,Aceves94,Capobianco95}

First, we will recall the principle used to estimate the stability of nonlinear modes using the criterion from Ref.~\cite{Mitchell93}. Then we will use it to analyze the stability of some of the plasmon--solitons found in NSWs.

\begin{figure}[!t]
	\centerline{\includegraphics[width=1\columnwidth,clip=true,trim= 0 0 0 0]{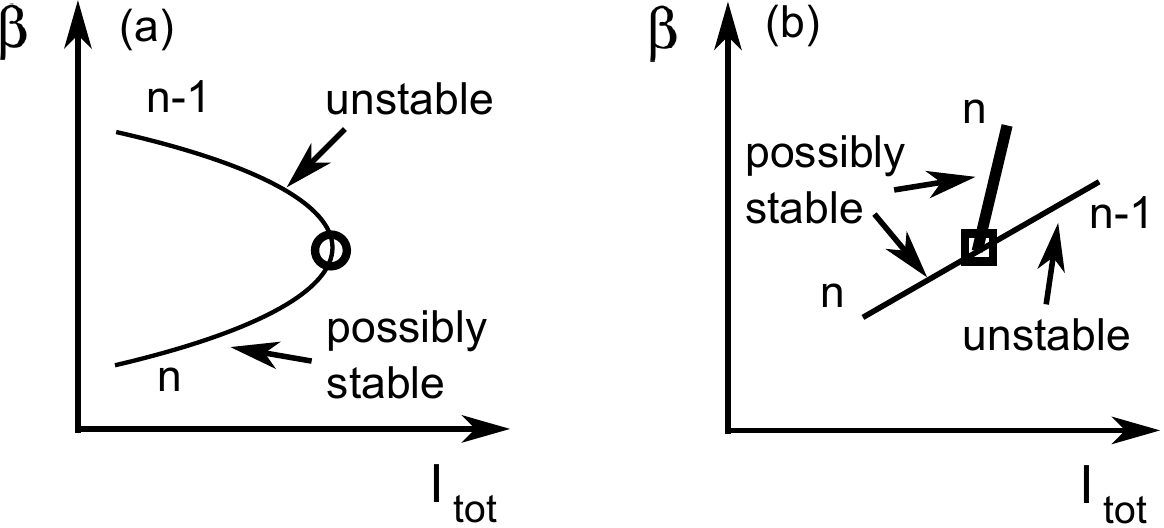}}
	\caption{Rules of assigning the stability of the modes for two specific cases extracted from Fig.~2 in Ref.~\cite{Mitchell93}: (a) the fold bifurcation (open circle) and (b) the Hopf bifurcation (open square). Thick lines indicate a doubly degenerate branch, whereas thin lines indicate non-degenerate dispersion curves.}
	\label{fig:stab-mitchell}
\end{figure}

The stability criterion derived in Ref.~\cite{Mitchell93} uses several assumptions. It provides stability for the fundamental nonlinear modes in structures composed of arbitrary nonlinear material distributed nonuniformly in the transverse direction. The derivation of the stability criterion from Ref.~\cite{Mitchell93} is obtained in the weak guiding approximation for which the electric field satisfies the scalar wave equation. In our study of the TM polarized waves, we consider the case in which it is the magnetic field component that satisfies the scalar wave equation [Eq.~(\ref{eqn:wave5}) in Ref.~\cite{Walasik14b}]. We are fully aware of the fact that the metal/nonlinear dielectric structures studied here, in which plasmon--soliton waves propagate, do not fulfill the weak guiding approximation due to high permittivity contrast between the metal and the nonlinear dielectric. This means that interesting nonlinear effects will occur for quite high nonlinear permittivity modifications. In spite of this fact, we use here the criterion from Ref.~\cite{Mitchell93}, because the dispersion diagrams obtained for our structures have similar character to the dispersion plots of the fully dielectric structures where the criterion is applicable and because, as it will be shown below, two different numerical propagation simulations of the full vector nonlinear problem confirm at least partially the theoretical predictions.

\begin{figure}[!t]
	\centerline{\includegraphics[width=1\columnwidth,clip=true,trim= 0 0 0 0]{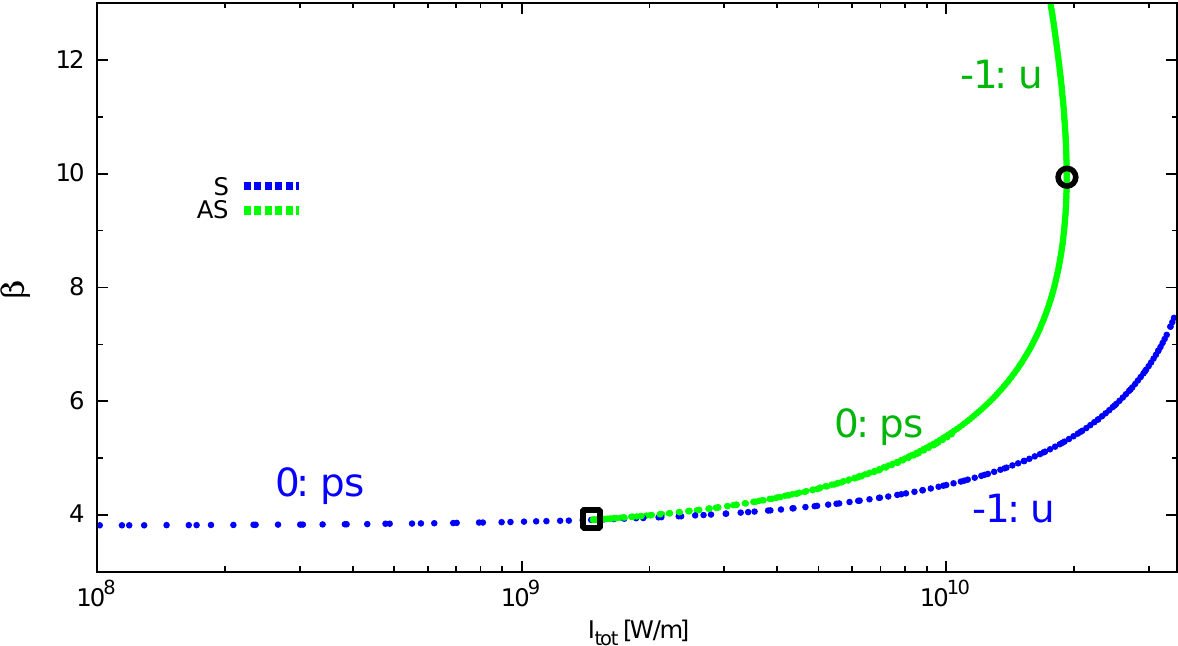}}
	\caption{Zoom on the region of the dispersion diagram with the birth of the first order asymmetric mode. Bifurcation points are marked with an open circle for fold bifurcation and an open square for Hopf bifurcation. The numbers facilitating the stability analysis are assigned to the sections of the dispersion curves according to the rules presented in Fig.~\ref{fig:stab-mitchell}. Labels 'ps' and 'u' denote possibly stable and unstable modes, respectively.}
	\label{fig:stab-low}
\end{figure}

In Fig.~\ref{fig:stab-mitchell}, the rules that will be required here to determine the stability of the modes derived in Ref.~\cite{Mitchell93} are schematically shown. Consider the dispersion relation presented in Fig.~\ref{fig:stab-low}. It shows  a zoom of a dispersion diagram, using $I_{\textrm{tot}}$ as variable, for a region that  contains  the dispersion curves of the main modes for the same structure as the one presented in Fig.~\ref{fig:disp-Pc}.
The stability of modes changes only at the bifurcation points~\cite{Mitchell93}. To determine the stability, first we have to identify all the bifurcation points on the dispersion diagram $\beta(I_{\textrm{tot}})$. In Fig.~\ref{fig:stab-low}, the bifurcation points are located at the points where intensity $I_{\textrm{tot}}$ has its local minima or maxima (point indicated by an open circle --- so-called fold bifurcation~\cite{Thompson02}) or where another branch appears [point indicated by an open square --- so called Hopf bifurcation associated with the birth of a doubly degenerate branch (to a single point on this branch correspond two asymmetric field profiles)]. Modes appear from or disappear at the points of bifurcation. The next step is to label the sections between the bifurcation points with numbers. The numbers are assigned in the following way. At first, we arbitrarily choose one section and label it with any number (in Fig.~\ref{fig:stab-low} we labeled the low intensity section of the symmetric dispersion curve with a number 0). The numbers of all the other sections of dispersion curves are assigned using the geometric rules given in Fig.~\ref{fig:stab-mitchell}.

Finally, after having numbered all the sections of the dispersion curves, we can read the stability of the modes directly from the $\beta(I_{\textrm{tot}})$ dispersion curves. The topological stability criterion presented in Ref.~\cite{Mitchell93} tells us that only the modes corresponding to the parts of the curves with the largest number are possibly stable. In Fig.~\ref{fig:stab-low}, only the modes labeled by 0 are possibly stable  (ps). All the other modes are unstable (u). The stability of all the possibly stable modes can be specified at once, as soon as the stability of one of them is determined. The stability can be determined either using numerical methods or theoretical arguments. The low-intensity section of the symmetric branch in the linear limit corresponds to a linear plasmon in metal/insulator/metal (MIM) configurations, which is stable. Therefore, the solutions corresponding to this section of the nonlinear dispersion curves should be stable. This hypothesis will be confirmed in Sec.~\ref{sec:stability-numerical}.

The high intensity section of the symmetric branch (above the Hopf bifurcation) corresponds to unstable solutions. On the contrary, the section of the asymmetric branch just above the Hopf bifurcation should correspond to stable solutions, because the stability properties of the sections with the same number are the same~\cite{Mitchell93}. 
On the asymmetric branch (at $\beta \approx 10$) another bifurcation occurs (fold type bifurcation indicated by an open circle). The high effective index section of the asymmetric branch (above the fold bifurcation point) is unstable.

\subsection{Numerical simulations of nonlinear propagations}
\label{sec:stability-numerical}

In the previous section we provide some results about the stability of the plasmon--solitons of the lowest orders using the topological criterion derived in Ref.~\cite{Mitchell93}. In the NSWs studied here the weak guiding approximation, used in the derivation of this topological criterion, is not fulfilled. This fact makes the conclusions drawn using the criterion not definitive. For this reason we also investigate the stability by full-vector numerical simulations.

\begin{figure}[!b]
  \centerline{\includegraphics[width=\columnwidth,height=0.21\columnwidth,clip=true,trim= 75 0 100 0]{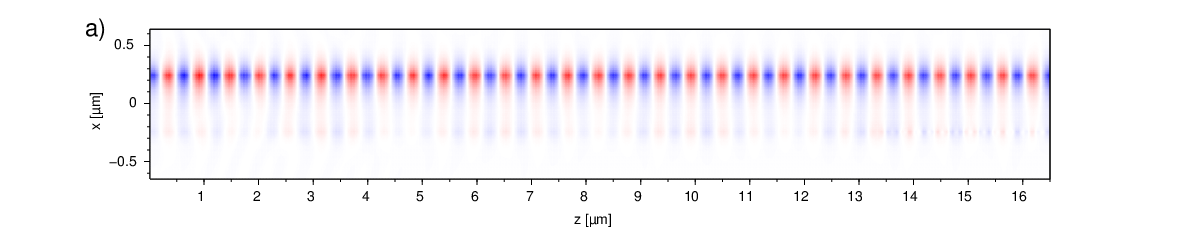}} 
  \centerline{\includegraphics[width=\columnwidth,clip=true,trim= 75 0 100 0]{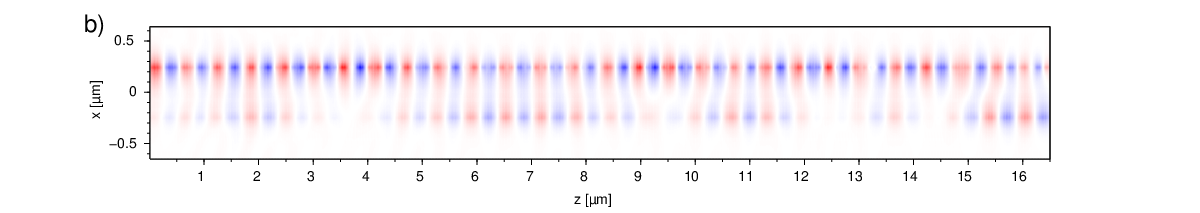}}
	\caption{Evolution of the $H_y$ field profile of an  asymmetric plasmon--soliton for the  stable case  (a) with   $\langle \Delta n \rangle = 0.0138$ and for the unstable case  (b) with  $\langle \Delta n \rangle = 0.005 $.  These simulations are realized using the FDTD method implemented in the \textsc{meep} software. The parameters are given in Fig.~\ref{fig:disp-stab-main-modes-fdtd}.}. 
	\label{fig:stab-asymmetric-mode-top-fdtd}
\end{figure}

First,  we have used the capabilities of the FDTD method~\cite{Taflove05,Taflove13} implemented in the \textsc{meep} software~\cite{meep-web,Oskooi10,Rodriguez07}. The metal permittivity is described by a Drude model to obtain the fixed negative value used at the studied wavelength in the semi-analytical models we developped. The useful computational domain is surrounded at its four edges by absorber regions that prevent back-reflected fields more efficiently than the perfectly matched layers that  have also been tested during our FDTD simulations.
An example of the FDTD propagation of the asymmetric plasmon--soliton is presented in Fig.~\ref{fig:stab-asymmetric-mode-top-fdtd}(a) through the evolution of the electric field component $H_y$, where the sinusoidal phase modifications are visible.

\begin{figure}[!b]
  \centerline{\includegraphics[width=\columnwidth,clip=true,trim= 75 0 100 0]{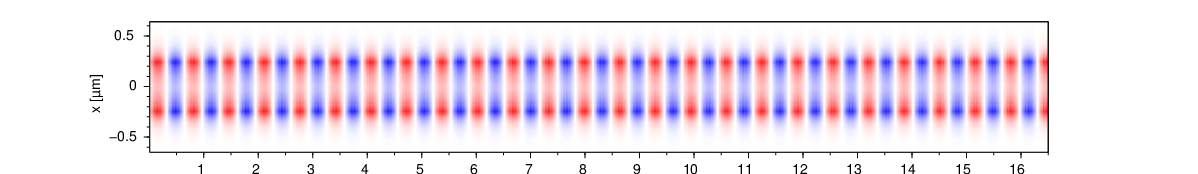}}
  \caption{Evolution of the $H_y$ field profile of a symmetric plasmon--soliton for the  stable case with   $\langle \Delta n \rangle = 0.0018$.  These simulations are realized using the FDTD method implemented in the \textsc{meep} software. The parameters are given in Fig.~\ref{fig:disp-stab-main-modes-fdtd}. } 
	\label{fig:stab-symmetric-mode-top-fdtd}
\end{figure}

 This result provides a confirmation of the stability of the first asymmetric mode in the nonlinear slot waveguides when the intensity is above a critical threshold. 
It can be noticed in Fig.~\ref{fig:stab-asymmetric-mode-top-fdtd}(a)  that the plasmon--soliton profile is not fully stationary. This is due to the fact that the used current excitations in our FDTD simulations do not generate perfectly the field profile of the asymmetric plasmon--soliton. The used asymmetric excitations contain components that weakly excite the antisymmetric plasmon--soliton (which is studied later in this section). 
When the excitations match perfectly with the asymmetric mode profile the observed non-stationary behaviour disappears as shown later when the simulation results from the other used  numerical method are described. The main symmetric plasmon-soliton is easier to excite in a simple way due to its symmetry property  as it can be seen in Fig.~\ref{fig:stab-symmetric-mode-top-fdtd} where a stable and stationary propagation is shown.

In order to obtain a more general view on the stability of the main modes of the NSW in the frame of the FDTD method, we systematically studied the propagation properties of the three main modes as a function of the spatially averaged refractive index variation $\langle \Delta n \rangle$  [see Eq.~(\ref{eq:definition-delta-n})]: the first symmetric, asymmetric, and antisymmetric modes.  
Typically, three cases occur in the simulation results:
\begin{itemize}
\item{Case 1: The mode is visible during the entire simulation duration. This case is, for example, the one encountered for the asymmetric mode above a given threshold $\langle \Delta n \rangle$ as it can be seen for example in Fig.~\ref{fig:stab-asymmetric-mode-top-fdtd}(a) or in Fig.~\ref{fig:stab-symmetric-mode-top-fdtd}.}
\item{Case 2: The studied mode is generated at the beginning of the temporal evolution but after some times it does not propagate anymore in a self-similar way. This case is the one encountered for the main asymmetric mode below a given  threshold $\langle \Delta n \rangle$ as shown  for example in Fig.~\ref{fig:stab-asymmetric-mode-top-fdtd}(b) where only the most stable part of the propagation is shown.} 
\item{Case 3: The investigated mode is not generated by the chosen current source (symmetric, antisymmetric or asymmetric) used to excite it,  even at the beginning of the temporal evolution and  in the surrounding of the source. This behaviour is observed for the asymmetric mode below the critical power or critical $\langle \Delta n \rangle$  associated to the Hopf bifurcation.} 
\end{itemize}
It is one of the main advantages of the FDTD method to be able to simulate temporal evolution even in the case of unstable modes unlike the other method used later in this section.

\begin{figure}[!b]
	\centerline{\includegraphics[width=\columnwidth,clip=true,trim= 0 0 0 0]{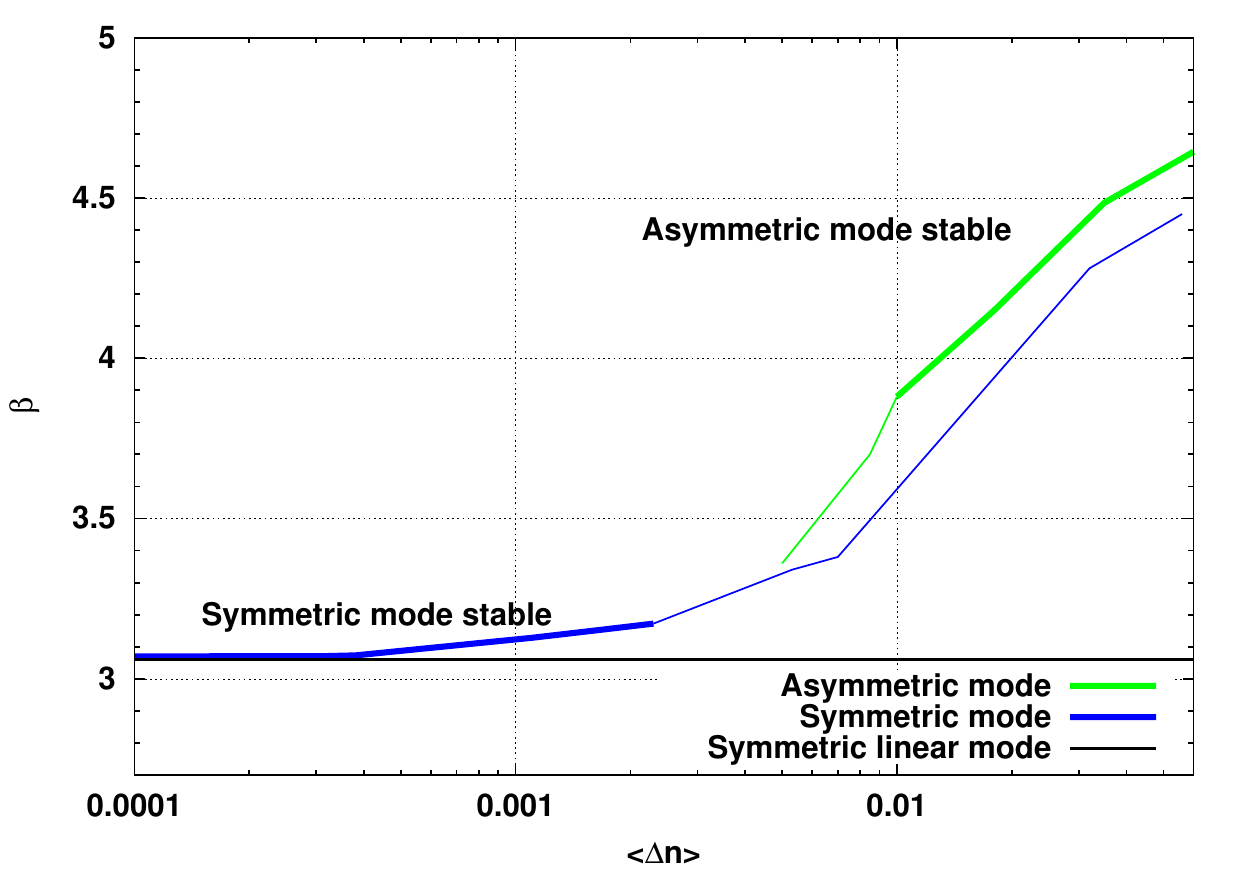}}
	\caption{Dispersion and stability results, obtained from FDTD simulations, for the first symmetric and asymmetric modes of the symmetric NSW. Thick curves denote a stable propagation while thin ones denote an unstable propagation.  The parameters are the  following: core permittivity $\epsilon_{l,2} = 3.46^2$, the second-order nonlinear refractive index $n_2^{(2)} = 2\cdot10^{-17}$~m$^2$/W, core thickness $d =500$~nm, metal permittivities $\epsilon_1 = \epsilon_3 = -6$ at a free-space wavelength $\lambda = 1.0$~$\mu$m.}
	\label{fig:disp-stab-main-modes-fdtd} 
\end{figure}

As  it is shown in Fig.~\ref{fig:disp-stab-main-modes-fdtd} obtained from the FDTD simulations, we are able to build a dispersion diagram for the first symmetric and asymmetric modes taking into account their stability properties. The given $\beta$ values for unstable modes, corresponding to the  case 2 in the above paragraph, are the ones extracted from the simulations results in the stable initial part of the evolution. It is evident that for the case 3 above, no dispersion data are obtained.

The stability properties of the asymmetric mode from the FDTD simulations differ from the ones deduced from the topological criterion  given in the previous section for  the stationary case. The asymmetric mode is not stable just above the bifurcation  (see  case 2  above) for some range of  $\langle \Delta n \rangle $  (see the thin green curve in Fig.~\ref{fig:disp-stab-main-modes-fdtd}), and then it becomes  stable  when  $\langle \Delta n \rangle $ increases (see the thick green curve in Fig.~\ref{fig:disp-stab-main-modes-fdtd}). 
The instability of the asymmetric mode just after the bifurcation has already been described in the field of the spatial soliton studies~\cite{Matuszewski07,Mayteevarunyoo08}. In our case, the instability can be observed in a relatively extended range of intensity or equivalently of $\langle \Delta n \rangle $.   
This extension of the instability could be due to the way the asymmetric mode is excited in our FDTD simulations and/or to the fact that the metal permittivity is dispersive due to the used Drude model.

It is worth noting that the FDTD dispersion curve for the asymmetric mode differs at high  $\langle \Delta n \rangle $ from the one computed  using the stationary IM: here the $\beta$ values are smaller and the FDTD curve stays concave while the stationary one is convex. Similar saturation effects in nonlinear full-vector temporal simulations  have already been described \textit{e.g.}, in Ref.~\cite{Akhmediev:93}.  From the FDTD implementation we use,  we can not conclude about the stability property at higher intensities than the ones shown in Fig.~\ref{fig:disp-stab-main-modes-fdtd} due to the limitations of the nonlinear treatment used (see reference~\cite{Oskooi10}).  Consequently, we can not check the stability properties around or above the fold bifurcation point described in Section~\ref{sec:stability-theoretical}.

As it was expected from the previous section, the first symmetric  mode is stable at low $\langle \Delta n \rangle $ or equivalently at low intensities (see the thick blue curve in Fig.~\ref{fig:disp-stab-main-modes-fdtd}). %

Its stability is lost slightly before the birth and the partial propagation of the asymmetric mode (see the  thin blue curve in Fig.~\ref{fig:disp-stab-main-modes-fdtd}). For all $\langle \Delta n \rangle $  values tested above this transition region, the first symmetric mode is unstable.
It is worth mentionning that the stability of this symmetric mode is recovered numerically as soon as the symmetry is forced in the FDTD simulations prohibiting the appearance of asymmetric behaviour.

\begin{figure}[!b]
	\centerline{\includegraphics[width=\columnwidth,clip=true,trim= 75 0 100 0]{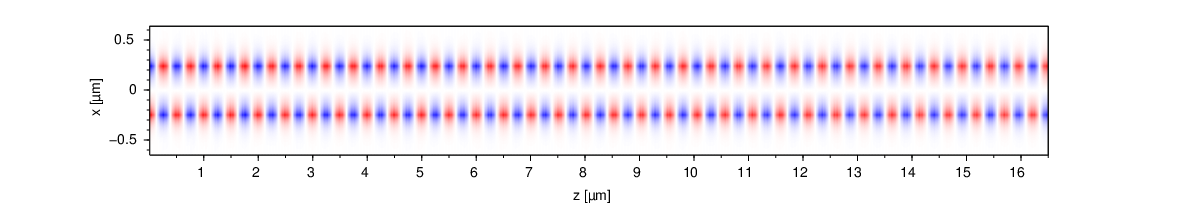}}
	\caption{Evolution of the $H_y$ field profile of a stable antisymmetric plasmon--soliton with $\langle \Delta n \rangle = 0.0225$.  These simulations are realized using the FDTD method implemented in the \textsc{meep} software.  The parameters are the same as in Fig.~\ref{fig:disp-stab-main-modes-fdtd}. } 
	\label{fig:stab-antisymmetric-mode-top-fdtd}
\end{figure}

The topological criterion given in given in Ref.~\cite{Mitchell93} can not be applied to the first antisymmetric plasmon--soliton mode because it is valid only for  fundamental modes. Therefore, the stability of this mode can only be infered from  numerical simulations.
The first antisymmetric mode starts, in the low-intensity regime, from the stable linear antisymmetric plasmon and there is no bifurcations on its dispersion curve. Therefore, we expect  this mode to be stable. An example of this stable propagation is shown in Fig.~\ref{fig:stab-antisymmetric-mode-top-fdtd}. We observe no change of the field profiles during the propagation (the antisymmetric excitation used does not contain any symmetric component).  

\begin{figure}[!h]
	\centerline{\includegraphics[width=\columnwidth,clip=true,trim= 0 0 0 0]{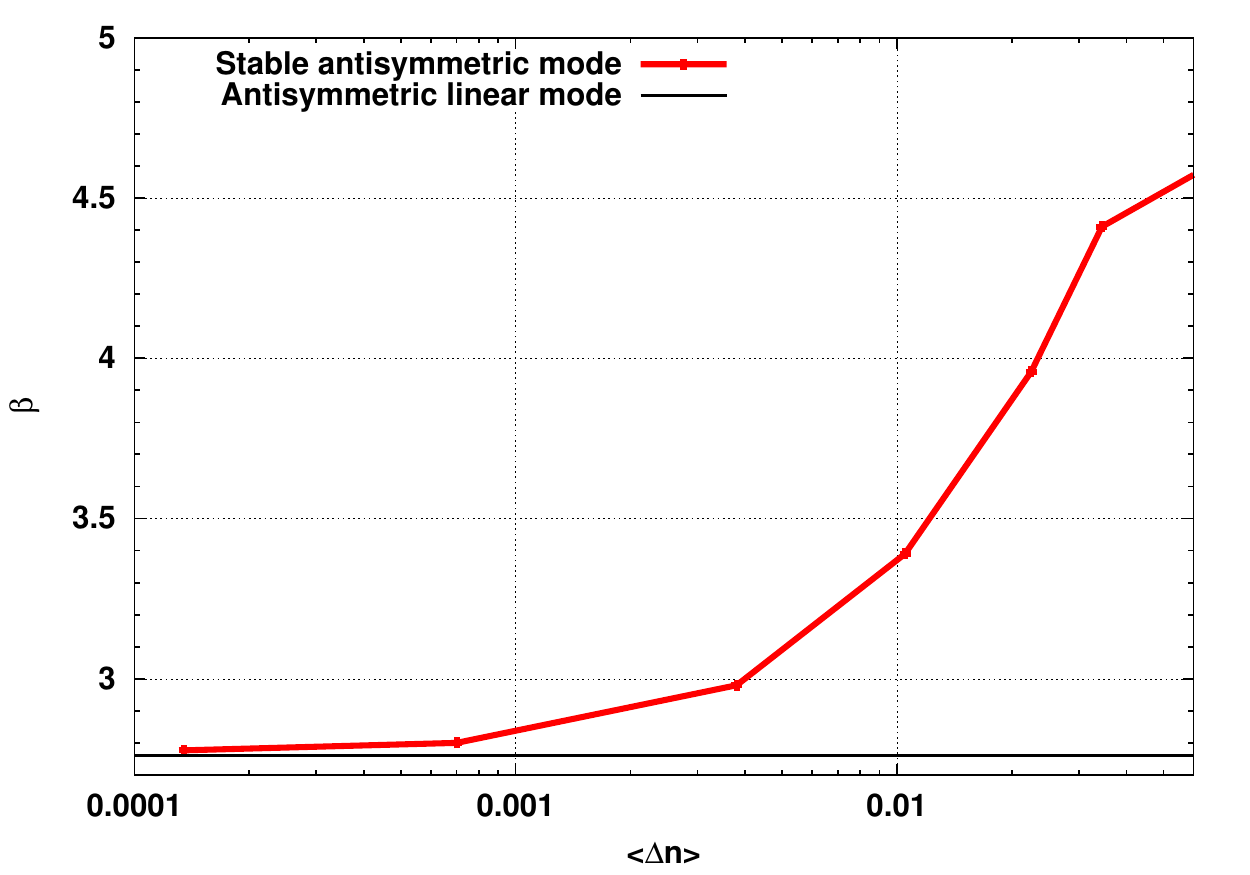}}
	\caption{Dispersion and stability results for the first antisymmetric  modes of the symmetric NSW obtained from FDTD simulations. The thick curve denotes a stable propagation.  The parameters are the same as in Fig.~\ref{fig:disp-stab-main-modes-fdtd}.}
	\label{fig:disp-stab-asymmetric-mode-fdtd} %
\end{figure}

In  Fig.~\ref{fig:disp-stab-asymmetric-mode-fdtd}, the dispersion curve of the antisymmetric plasmon--soliton is given together with its stability property. The antisymmetric mode is stable up to the maximum intensity that can be treated with the FDTD implementation we use.

The stability properties of the three main plasmon--soliton modes in nonlinear slot waveguides are also verified using the nonlinear propagation scheme implemented in the lastest version of the RF module of COMSOL Multiphysics~\cite{comsol-web}. This approach was successfully used to study the stability of solitons in lattices built of metals and nonlinear dielectrics~\cite{Kou13,Huang13,Shi14}. This method is limited to the cases where the studied mode is stable since the iterative numerical method used to compute the fields do not converge in other cases. 

\begin{figure}[!h]
	\centerline{\includegraphics[width=0.55\columnwidth,clip=true,trim= 0 0 0 0]{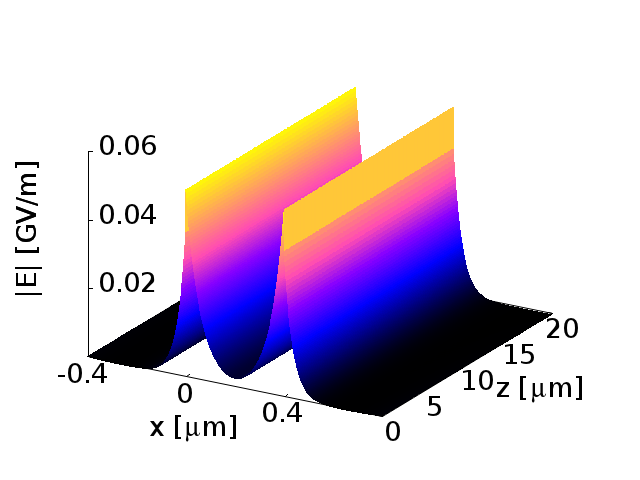}}
	\caption{Evolution of the electric field norm during the propagation of the symmetric mode located below the Hopf bifurcation threshold. The average nonlinear index change in the core induced by this mode is equal to $\langle \Delta n \rangle = 10^{-4}$ and the propagation distance is approximately 13 free-space wavelengths.  The parameters are the  following: core permittivity $\epsilon_{l,2} = 3.46^2$, the second-order nonlinear refractive index $n_2^{(2)} = 2\cdot10^{-17}$~m$^2$/W, core thickness $d =400$~nm, metal permittivities $\epsilon_1 = \epsilon_3 = -20$ at a free-space wavelength $\lambda = 1.55$~$\mu$m. These simulations are realized using the COMSOL software.}
	\label{fig:stab-sym-comsol}
\end{figure}

\begin{figure}[!h]
  \centerline{\includegraphics[width=0.5\columnwidth,clip=true,trim= 0 0 0 0]{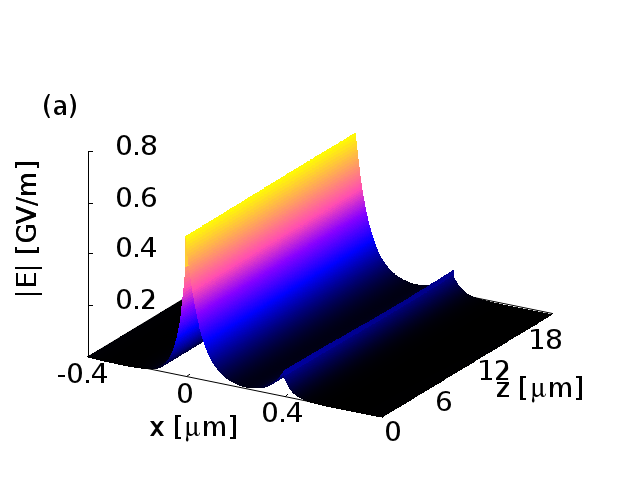}
    \includegraphics[width=0.5\columnwidth,clip=true,trim= 0 0 0 0]{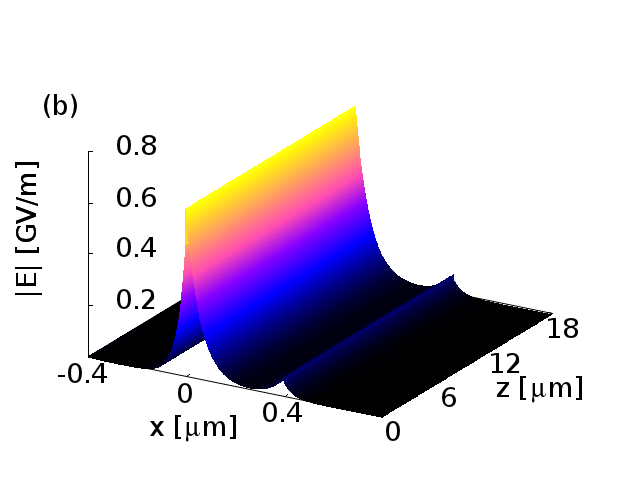}}
  \centerline{\includegraphics[width=0.5\columnwidth,clip=true,trim= 0 0 0 0]{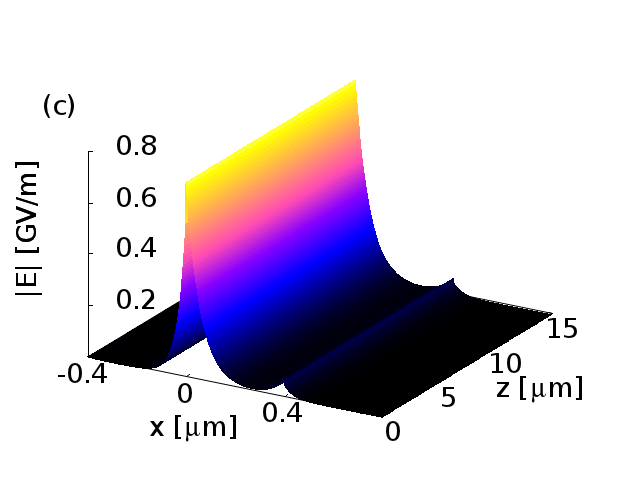}}
  \caption{Evolution of the electric field norm during the propagation of asymmetric modes located between the Hopf bifurcation and the fold bifurcation. The average nonlinear index change in the core $\langle \Delta n \rangle$ induced by these modes is equal to (a) $2 \cdot 10^{-3}$, (b) $3 \cdot 10^{-3}$, and (c) $4 \cdot 10^{-3}$. The shown propagation distance is approximately 12 free-space wavelengths. The parameters are the same as in Fig.~\ref{fig:stab-sym-comsol}. These simulations are realized using the COMSOL software.}
	\label{fig:stab-asym-comsol}
\end{figure}

\begin{figure}[!t]
	\centerline{\includegraphics[width=\columnwidth,clip=true,trim= 20 0 20 0]{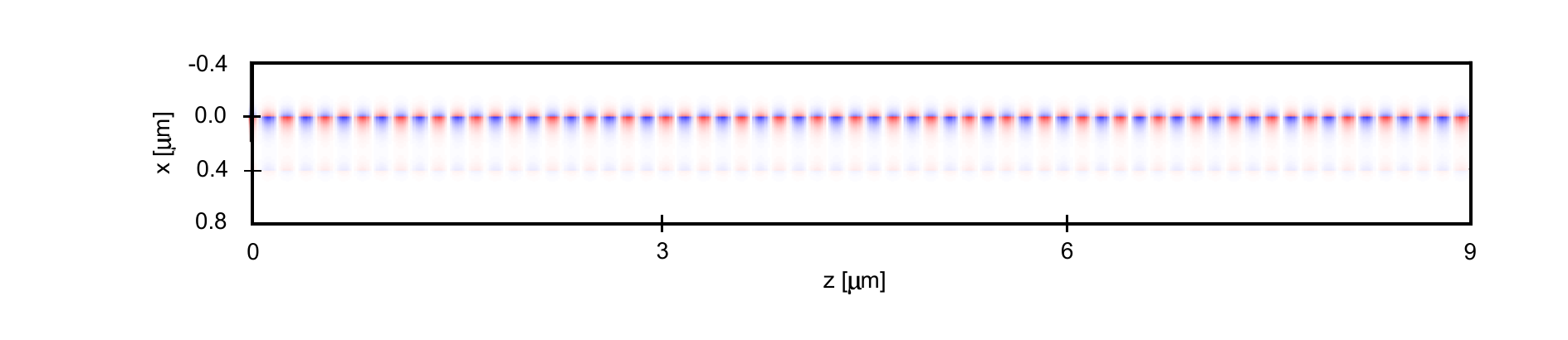}}
	\caption{Evolution of the $E_x$ field profile during the propagation of the solution presented in Fig.~\ref{fig:stab-asym-comsol}(b) for a slot with $d= 400$~nm. These simulations are realized using the COMSOL software. The shown propagation distance is approximately 6 free-space  wavelengths. }
	\label{fig:stab-asym-top-comsol} 
\end{figure}

According to the conclusions drawn from  Fig.~\ref{fig:stab-low} in Sec.~\ref{sec:stability-theoretical} and from the FDTD simulations, the low-power section of the symmetric branch corresponds to stable solutions. This result is also  confirmed by the simulation presented in Fig.~\ref{fig:stab-sym-comsol} obtained for a NSW with $d=400$~nm. %
No stable symmetric solution was found above  the $ \langle \Delta n \rangle $ transition region  using the  method implemented in COMSOL confirming  the FDTD results already obtained.
 The stability of the asymmetric branch above the bifurcation region observed in Fig.~\ref{fig:disp-stab-main-modes-fdtd}  is confirmed by these numerical simulations as shown in Fig.~\ref{fig:stab-asym-comsol}. Figure~\ref{fig:stab-asym-top-comsol} presents the evolution of the $E_x$  electric field component  for the asymmetric solutions in such a case.%

\begin{figure}[!h]
	\centerline{\includegraphics[width=0.5\columnwidth,clip=true,trim= 0 0 0 0]{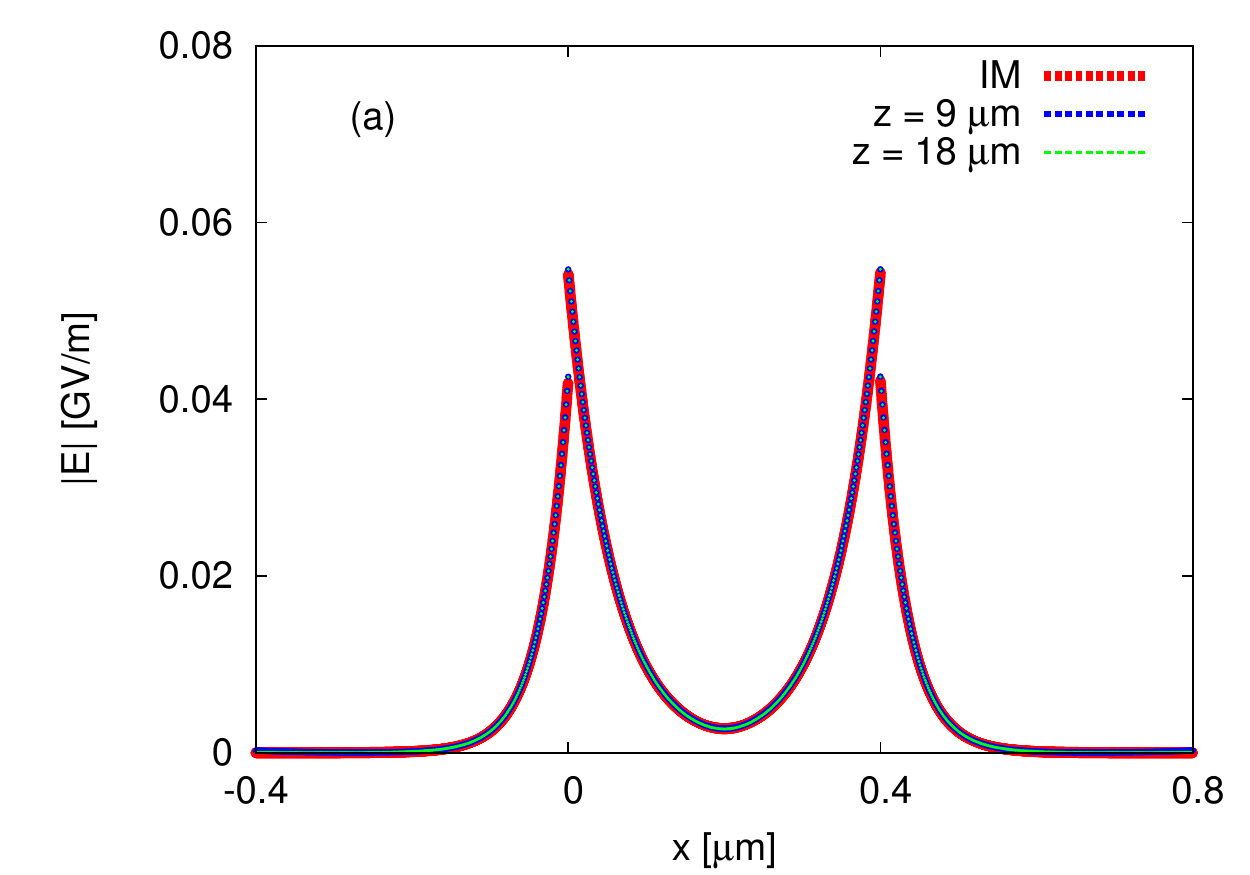}
		\includegraphics[width=0.5\columnwidth,clip=true,trim= 0 0 0 0]{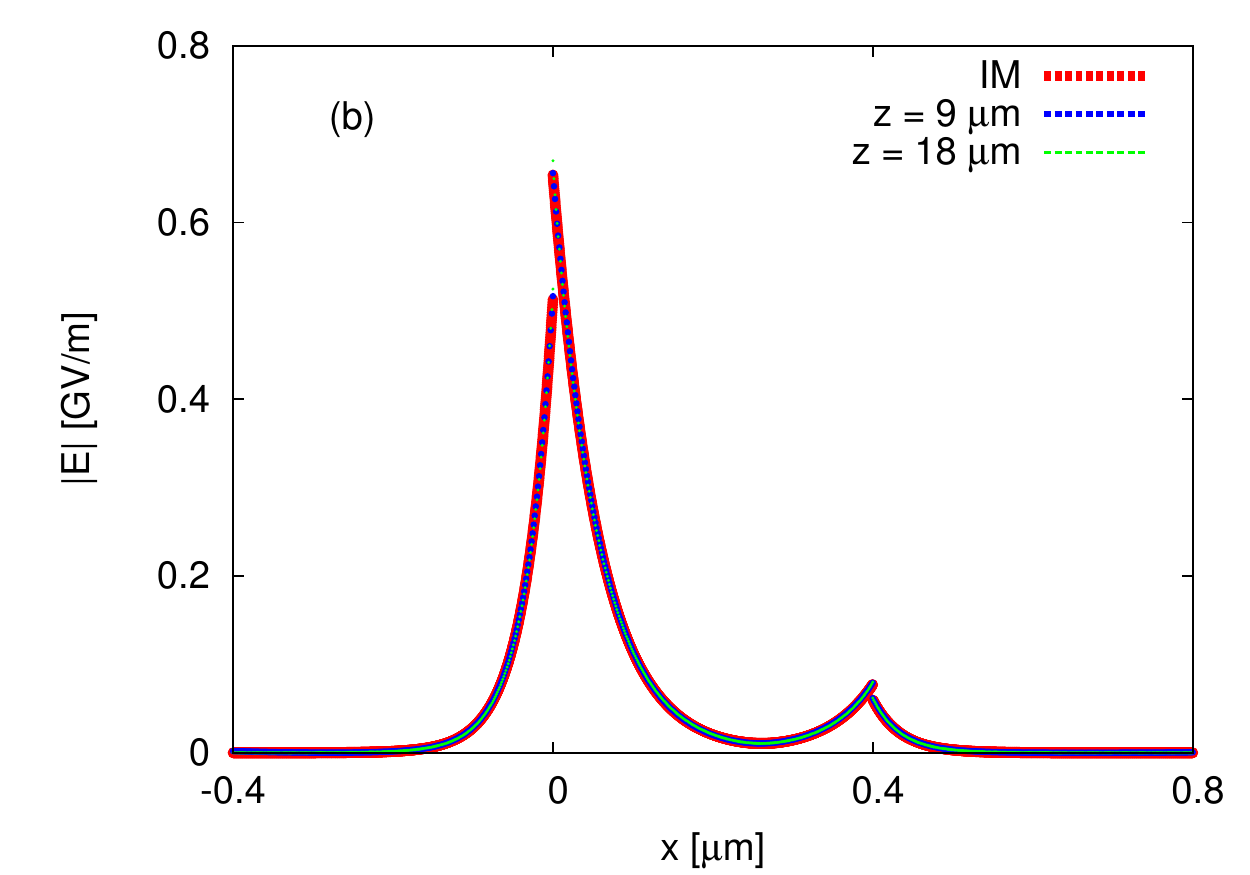}}
	\caption{Comparison of the $|\textbf{E}|$ profiles obtained using the IM (and used as the input profiles in the COMSOL based simulations) and cuts of the field evolution in the middle of the propagation range ($z = 9$~$\mu$m --- 6 free-space wavelengths) and at the end of the propagation ($z = 18$~$\mu$m ---  12 free-space wavelengths) for (a) the symmetric nonlinear plasmon--soliton (see Fig.~\ref{fig:stab-sym-comsol}) and (b) the asymmetric nonlinear plasmon--soliton [see Fig.~\ref{fig:stab-asym-comsol}(b)].}
	\label{fig:stab-cut-comsol-compar}
\end{figure}

Figure~\ref{fig:stab-cut-comsol-compar} shows the transverse profiles of the symmetric and asymmetric plasmons--solitons in the NSW. For each symmetry type, we compare the profiles obtained using the interface model (these profiles are used as input in the COMSOL based propagation simulations) with the cuts of the profiles presented in Figs.~\ref{fig:stab-sym-comsol} and \ref{fig:stab-asym-comsol}(b). 
These comparisons validate the accuracy of the evolution simulations and consequently the results obtained concerning the stability properties of the symmetric and asymmetric modes in the NSW. One can notice that the stationary behaviour is more clearly seen in the COMSOL based simulations than in the FDTD ones. This is due to the fact that in the former case we directly use as input the profiles provided by the interface model while in the latter case we use excitation current sources  to generate the fields that mimick the stationary field profiles. Since we are investigating nonlinear phenomena, it is not possible to use in the FDTD simulations  a part of a linear waveguide to filter the needed profile as it is usually  done in FDTD based linear studies~\cite{Taflove05}.

\section{Conclusions}
We have provided detailed results for the plasmon-soliton waves in planar slot waveguides with a finite-thicknees nonlinear dielectric core. In symmetric structures, using the semi-analytical models we developped for stationary states, we haved investigated the properties of the first main modes and reported new higher order modes including asymmetric ones that exist at high intensities only. We have also described complete dispersion diagrams for these different modes as a function of various quantities including the total power, the field value at one interface between the metal and the nonlinear core, and also the spatial average of the nonlinear refractive index change. We have proven that the total intensity or equivalently the spatially averaged nonlinear refractive index change corresponding to the Hopf bifurcation threshold  from the first symmetric mode to the first asymmetric mode can be reduced by several orders of magnitude with an increase of the permittivity of the core or of the metal cladding. 
We have also proven the versatility of our semi-analytical models studying asymmetric structures. For such structures, we have described the impact of the metal permittivity contrast that lifts the degeneracy of the doubly degenerated asymmetric mode providing a more complex dispersion diagram than the one of a symmetric structure.

Concerning, the stability of the main symmetric and asymmetric modes, we have used  an already derived topological criterion established only in the weak guidance approximation being fully aware that our structures lay beyond its validity range. Nonetheless, as shown by full-vector simulations, the topological criterion predicts correctly the principal stability properties of the main modes of the studied planar nonlinear slot waveguides.  Using this criterion, we have shown that the asymmetric mode emerging through a Hopf bifurcation at a critical intensity is stable  between  this bifurcation and a fold bifurcation located at higher intensity level. The stability of this asymmetric mode is lost at this fold bifurcation. On the contrary, the symmetric mode is unstable for all intensity levels above the Hopf bifurcation while it is stable below.

Using two different full-vector numerical propagation methods, we have studied the stability of the three main modes: the symmetric,  asymmetric, and antisymmetric modes.
 We have shown that the asymmetric mode  is stable above a critical intensity slightly larger than the threshold associated with the Hopf bifurcation computed for the stationary states from our semi-analytical models at least up to the maximum level of tested intensities.  The symmetric mode is shown to be unstable slightly below and slightly above  the Hopf bifurcation threshold, and to be stable at lower intensities. For all tested intensities, these results confirm qualitatively the results derived from the  topological criterion  even if quantitative differences exist.
Finally, we have also proven numerically that the anti-symmetric mode is stable in  the entire range of tested intensities.

These stability results together with those about the decrease of the bifurcation threshold should facilitate the design of specific structures in order to make possible the experimental observation of these plasmon-soliton waves more than thirty years after their theoretical discovery. Future studies should be dedicated to the further reduction of the bifurcation threshold and to the study of more sophisticated configurations.

\vspace{0.25cm}

\begin{acknowledgments}
This work was supported by the European Commission through the Erasmus Mundus Joint Doctorate Programme  Europhotonics (Grant No. 159224-1-2009-1-FR-ERA MUNDUS-EMJD). This work was not funded by the French Agence Nationale de la Recherche. W.~W and G.~R. would like to thank Yaroslav Kartashov for helpful comments on this work and  Alejandro Rodriguez for useful discussions about the \textsc{meep} software. G.~R. would like to thank the Scilab consortium.
\end{acknowledgments}

\bibliographystyle{apsrev4-1}
\bibliography{PRA-14-part2}{}

\end{document}